\documentstyle[epsfig, natbib209]{mn}

\newif\ifAMStwofonts

\ifoldfss
  \newcommand{\rmn}[1] {{\rm #1}}
  \newcommand{\itl}[1] {{\it #1}}

  \ifCUPmtlplainloaded \else
    \NewTextAlphabet{textbfit} {cmbxti10} {}
    \NewTextAlphabet{textbfss} {cmssbx10} {}
    \NewMathAlphabet{mathbfit} {cmbxti10} {} 
    \NewMathAlphabet{mathbfss} {cmssbx10} {} 
  \fi
  \ifAMStwofonts
    \ifCUPmtlplainloaded \else
      \NewSymbolFont{upmath} {eurm10}
      \NewSymbolFont{AMSa} {msam10}
      \NewMathSymbol{\upi}     {0}{upmath}{19}
      \NewMathSymbol{\umu}     {0}{upmath}{16}
      \NewMathSymbol{\upartial}{0}{upmath}{40}
      \NewMathSymbol{\leqslant}{3}{AMSa}{36}
      \NewMathSymbol{\geqslant}{3}{AMSa}{3E}

      \let\leq=\leqslant \let\le=\leqslant
      \let\geq=\geqslant \let\ge=\geqslant
    \fi
  \fi
\fi 

\ifnfssone
  \newmathalphabet{\mathit}
  \addtoversion{normal}{\mathit}{cmr}{m}{it}
  \addtoversion{bold}{\mathit}{cmr}{bx}{it}
  \newcommand{\rmn}[1] {\mathrm{#1}}
  \newcommand{\itl}[1] {\mathit{#1}}

  \newmathalphabet{\mathbfit} 
  \addtoversion{normal}{\mathbfit}{cmr}{bx}{it}
  \addtoversion{bold}{\mathbfit}{cmr}{bx}{it}
  \newmathalphabet{\mathbfss} 
  \addtoversion{normal}{\mathbfss}{cmss}{bx}{n}
  \addtoversion{bold}{\mathbfss}{cmss}{bx}{n}
  \ifAMStwofonts
    \ifCUPmtlplainloaded \else
      \UseAMStwoboldmath
      \makeatletter
      \new@@mathgroup\upmath@@group
      \define@@mathgroup\mv@@normal\upmath@@group{eur}{m}{n}
      \define@@mathgroup\mv@@bold\upmath@@group{eur}{b}{n}
      \edef\UPM{\hexnumber\upmath@@group}
      \new@@mathgroup\amsa@@group
      \define@@mathgroup\mv@@normal\amsa@@group{msa}{m}{n}
      \define@@mathgroup\mv@@bold\amsa@@group{msa}{m}{n}
      \edef\AMSa{\hexnumber\amsa@@group}
      \makeatother
      \mathchardef\upi="0\UPM19
      \mathchardef\umu="0\UPM16
      \mathchardef\upartial="0\UPM40
      \mathchardef\leqslant="3\AMSa36
      \mathchardef\geqslant="3\AMSa3E

      \let\leq=\leqslant \let\le=\leqslant
      \let\geq=\geqslant \let\ge=\geqslant
    \fi
  \fi
\fi 

\ifnfsstwo
  \newcommand{\rmn}[1] {\mathrm{#1}}
  \newcommand{\itl}[1] {\mathit{#1}}

  \DeclareMathAlphabet{\mathbfit}{OT1}{cmr}{bx}{it}
  \SetMathAlphabet\mathbfit{bold}{OT1}{cmr}{bx}{it}
  \DeclareMathAlphabet{\mathbfss}{OT1}{cmss}{bx}{n}
  \SetMathAlphabet\mathbfss{bold}{OT1}{cmss}{bx}{n}
  \ifAMStwofonts
    \ifCUPmtlplainloaded \else
      \DeclareSymbolFont{UPM}{U}{eur}{m}{n}
      \SetSymbolFont{UPM}{bold}{U}{eur}{b}{n}
      \DeclareSymbolFont{AMSa}{U}{msa}{m}{n}
      \DeclareMathSymbol{\upi}{0}{UPM}{"19}
      \DeclareMathSymbol{\umu}{0}{UPM}{"16}
      \DeclareMathSymbol{\upartial}{0}{UPM}{"40}
      \DeclareMathSymbol{\leqslant}{3}{AMSa}{"36}
      \DeclareMathSymbol{\geqslant}{3}{AMSa}{"3E}

      \let\leq=\leqslant \let\le=\leqslant
      \let\geq=\geqslant \let\ge=\geqslant
    \fi
  \fi
\fi 

\ifCUPmtlplainloaded \else
  \ifAMStwofonts \else 
    \def\upi{\pi}
    \def\umu{\mu}
    \def\upartial{\partial}
  \fi
\fi

\title{The first stars: formation of binaries and small multiple systems}
\author[A. Stacy, T. H. Greif, and V. Bromm]
       {Athena Stacy$^{1,2}$\thanks{E-mail: minerva@astro.as.utexas.edu}, Thomas H. Greif $^{3,4}$ and Volker Bromm$^{1,2}$\\
 $^{1}$Department of Astronomy, University of Texas, Austin, TX 78712, USA \\
 $^{2}$Texas Cosmology Center, University of Texas, Austin, TX 78712, USA \\
$^{3}$Institut f\"{u}r theoretische Astrophysik, Albert-Ueberle Strasse 2, 69120 Heidelberg, Germany\\
$^{4}$Fellow of the International Max Planck Research School for Astronomy and Cosmic Physics at the University of Heidelberg}

\pagerange{\pageref{firstpage}--\pageref{lastpage}}
\pubyear{2009}

\begin{document}

\maketitle
\topmargin-1cm

\label{firstpage}

\begin{abstract}
We investigate the formation of metal-free, Population III (Pop~III), stars within a minihalo
at $z\simeq 20$ with a smoothed particle hydrodynamics (SPH) simulation,
starting from cosmological initial conditions. Employing a hierarchical,
zoom-in procedure, we achieve sufficient numerical resolution to follow
the collapsing gas in the center of the minihalo up to number densities
of $10^{12}$\,cm$^{-3}$. This allows us to study
the protostellar accretion onto the initial hydrostatic core, which
we represent as a growing sink particle,
in improved physical detail. The accretion process, and in particular its
termination, governs the final masses that were reached by the first stars.
The primordial initial mass function (IMF),
in turn, played an important role in determining to what extent the first stars drove early cosmic evolution. We continue our simulation for 5000\,yr after the first sink particle has formed. During this time period, a disk-like configuration is assembled around the first protostar. The disk is gravitationally unstable, develops a pronounced spiral structure, and fragments into several other protostellar seeds. At the end of the simulation, a small multiple system has formed, dominated by a binary with masses $\sim 40$\,M$_{\odot}$ and $\sim 10$\,M$_{\odot}$. If Pop~III stars were
to form typically in binaries or small multiples, 
the standard model of primordial star formation,
where single, isolated stars are predicted to form in minihaloes, would have
to be modified. This would have crucial consequences for the observational
signature of the first stars, such as their nucleosynthetic pattern, and
the gravitational-wave emission from possible Pop~III black-hole binaries.
\end{abstract}


\section{Introduction}
How did the cosmic dark ages end, and how was the homogeneous early
Universe transformed into the highly complex state that we observe
today  (e.g. \citealt{barkana&loeb2001,miralda-escude2003,ciardi&ferrara2005})?
One of the key questions is to understand the formation and properties of the first stars, the so-called Population~III (or Pop~III), since they likely played a crucial role in driving early cosmic evolution  (e.g. \citealt{bromm&larson2004,glover2005,byhm2009}). For instance, the radiation from the first stars contributed to the reionization of the intergalactic medium (IGM), at least during its initial phase (e.g. \citealt{kitayamaetal2004,syahs2004,whalenetal2004,alvarezetal2006, johnsongreif&bromm2007}). After the first stars exploded as supernovae (SNe) they spread heavy elements through the IGM, thereby providing the initial cosmic metal enrichment (e.g. \citealt{madauferrara&rees2001,moriferrara&madau2002,brommyoshida&hernquist2003,wada&venkatesan2003,normanetal2004,greifetal2007,tfs2007}). The resulting metallicity was likely crucial in governing the transition from early high-mass star formation to the normal-mass star formation seen today (e.g. \citealt{omukai2000,brommetal2001,schneideretal2002,bromm&loeb2003,frebeletal2007,frebeletal2009,smith&sigurdsson2007}).

The extent to which the radiation and metal enrichment from the first stars influenced the early Universe is largely determined by their mass.  Currently, Pop III stars are believed to have first formed around $z\sim 20$ within minihaloes of mass $M\sim 10^6$\,M$_{\odot}$ (e.g. \citealt{haimanetal1996,tegmarketal1997,yahs2003}). The potential wells of these minihaloes are provided by dark matter (DM), which serve to initially gather primordial gas into them.  This is in contrast to present-day star-forming giant molecular clouds (GMCs), which are of similar mass to minihaloes, but which form within much larger galaxies and are not DM dominated. Previous three-dimensional simulations (e.g. \citealt{abeletal2002,brommetal2002}) have studied the formation of $M\sim 1000$\,M$_{\odot}$ prestellar cores with sizes $<$ 0.5 pc within such minihaloes. The density in these cores is similar to those in present-day ones, but having considerably higher temperature ($\sim 200$\,K for primordial cores versus $10 - 15$\,K for those within GMCs).  

The mass of the first stars depends on the further evolution of such cores. Similar to present-day star formation (e.g. \citealt{mckee&tan2002}), it is expected that some of the gas within these cores will gravitationally collapse into one or more protostars of initial mass $\sim 5 \times 10^{-3}$ M$_{\odot}$ (\citealt{omukai&nishi1998,yoh2008}). A fraction of the core mass will be accreted onto the protostar(s), ultimately forming stars much more massive than the initial seeds. Three-dimensional simulations which studied the growth of a primordial protostar through accretion imply that the first stars were quite massive, finding an upper limit of $\sim 500$\,M$_{\odot}$ (\citealt{bromm&loeb2004}). Some of the earliest studies of Pop III star formation, however, predicted that Pop III stellar masses might be quite low, $<$ 1 M$_{\odot}$ (e.g. \citealt{Kashlinsky1983}). These authors emphasized the importance of angular momentum in determining the mass of Pop III stars, predicting that rotational effects would cause the primordial gas clouds to collapse into a dense disk.  Only after the disk cooled to $\sim 1000$\,K through H$_2$ line emission could fragmentation occur. Depending on the spin of the cloud, the fragments thus formed could have had masses as low as $\la$ 0.2~M$_{\odot}$.  Later studies by \cite{saigoetal2004} also found that high initial angular momentum would allow primordial clouds to form disks, and they further predicted that disk fragmention into binaries would be common.  Similar calculations by \cite{machidaetal2008}, which were initialized at much lower densities and therefore at an earlier point in the gas collapse, found fragmentation would occur even for primordial gas clouds with very modest initial angular momentum.  
Finally, the recent work by \cite{clarketal2008} included a simulation of a 500 M$_{\odot}$ primordial gas cloud which fragmented into $\sim$ 20 stellar objects.
However, none of the above-mentioned investigations were initialized on cosmological scales.    

Our goal in this study is to further improve our understanding of how a Pop~III star is assembled through the process of accretion within the first minihaloes. More specifically, we aim to determine the final masses reached by the first stars, and whether the gas within a host minihalo forms a single star or fragments into a binary or small multiple system. To this end we perform a three-dimensional cosmological simulation that follows the formation of a minihalo with very high numerical resolution, allowing us to trace the evolution of its gas up to densities of  $10^{12}$\,cm$^{-3}$. Unlike the study of \citet{bromm&loeb2004}, which employed a constrained initialization technique on scales of $\sim$ 150\,pc, our simulation is initialized on cosmological scales. This allows the angular momentum of the minihalo to be found self-consistently, eliminating the need for an assumed spin parameter. In a fashion similar to \citet{bromm&loeb2004}, we represent high-density peaks that are undergoing gravitational runaway collapse 
using the sink particle method first introduced by \cite{bateetal1995}, 
allowing us to follow the mass flow onto the sinks well beyond the onset of collapse.  

The recent simulation by \citet{turketal2009}, starting from cosmological initial conditions, has shown fragmentation and the formation of two high-density peaks, which are interpreted as seeds for a Pop~III binary. These peaks are similar to the multiple sinks found in our simulation. Although Turk et al. employed even higher numerical resolution, they were able to follow the evolution of the high-density peaks for only $\sim$ 200 yr. Our sink particle method, on the other hand, enabled us to continue our calculation for several thousand years after initial runaway collapse. We could thus follow the subsequent disk evolution and further fragmentation in detail, rendering our study ideally complementary to their work.

The structure of our paper is as follows. In Section~2, we discuss the basic
physical ingredients determining the accretion flow, whereas we describe our
numerical methodology in Section~3. We proceed to present our results (Section~4), and conclude in Section~5.

\begin{figure*}
\includegraphics[width=.7\textwidth]{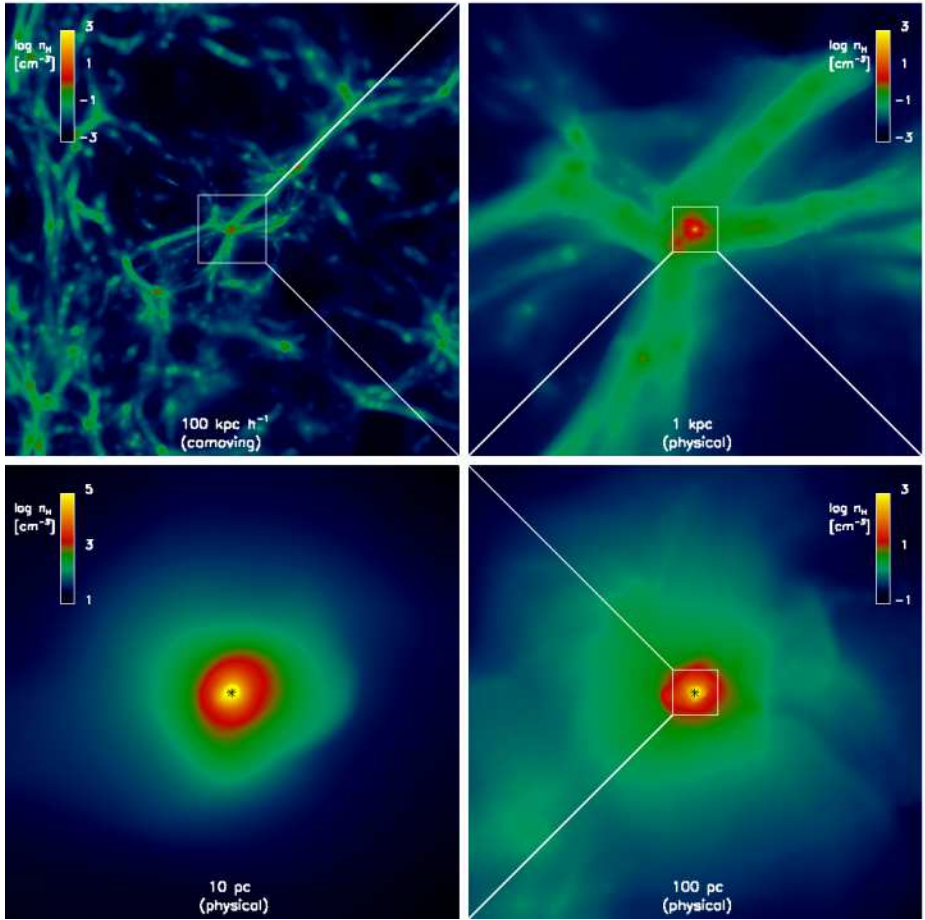}
\caption{Final simulation output shown in density projection along the x-z plane on progressively smaller scales, as labeled on each panel.  White boxes denote the region to be depicted in the next-smallest scale.  In the bottom two panels, the asterisk denotes the location of the first sink formed.  {\it Top left:} Entire simulation box.  {\it Top right:} Minihalo and surrounding filamentary structure.  {\it Bottom right:} Central 100 pc of minihalo.  {\it Bottom left:} Central 10 pc of minihalo. Note how the morphology approaches an increasingly
smooth, roughly spherical distribution on the smallest scales, where the
gas is in quasi-hydrostatic equilibrium, just prior to the onset of
runaway collapse.
}
\label{zoom_in}
\end{figure*}

\begin{figure*}
\includegraphics[width=.8\textwidth]{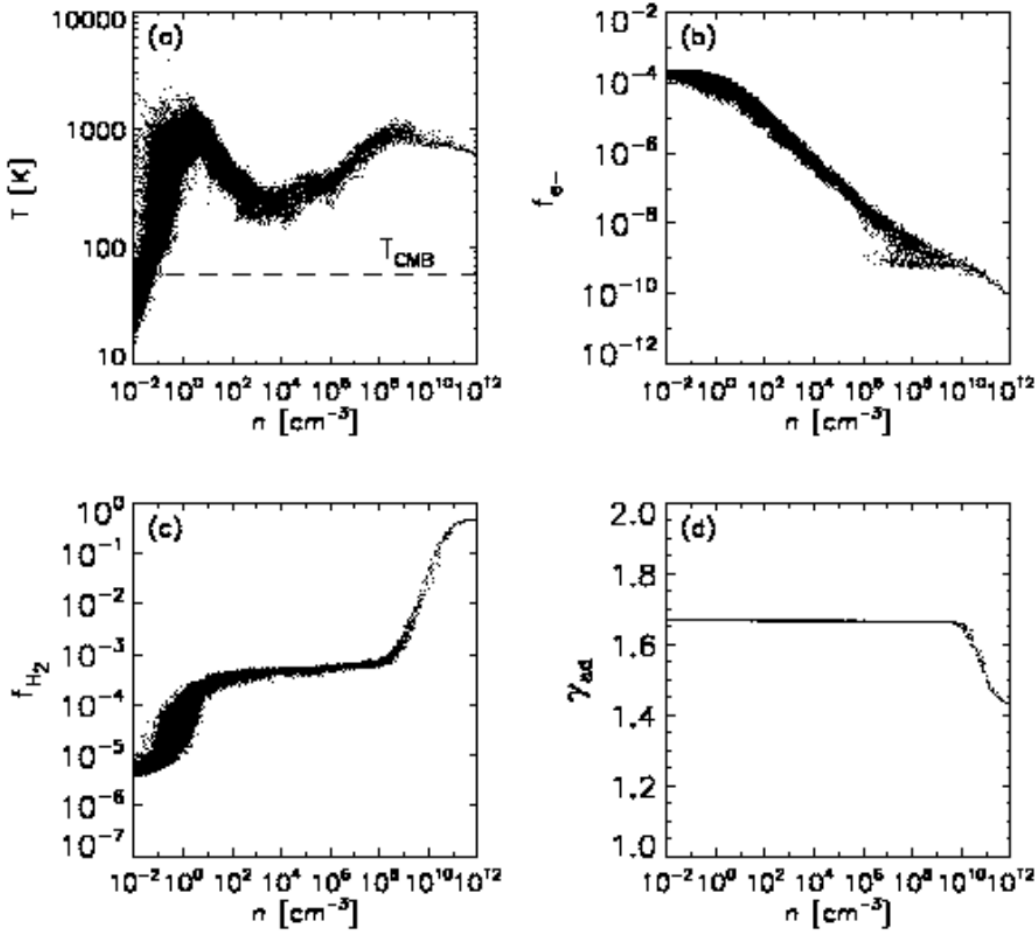}
\caption{
Physical state of the collapsing core. The situation is shown just prior to
the formation of the first sink particle.
{\it (a)} Temperature; {\it (b)} free electron fraction;
{\it (c)} molecular hydrogen fraction; and
{\it (d)} adiabatic exponent $\gamma_{\rmn ad}$ vs. density.
The gas that collapses into the
minihalo is heated adiabatically to $\sim 1000$\,K as it reaches  $n\sim 1$
cm$^{-3}$ (panel a). At this point $f_{\rmn H_{2}}$ becomes
high enough (panel c) to allow the gas to cool through H$_{2}$ rovibrational
transitions to a minimum of $T \sim 200$\,K, which is reached at $n \sim$
10$^4$ cm$^{-3}$ (panel a). After the onset of gravitational instability, the gas reaches yet higher densities, and is heated to a
maximum of $\sim$ 1000\,K once it approaches
a density of $\sim$ 10$^8$\,cm$^{-3}$. At this density three-body reactions
become important, and the gas is rapidly converted into fully molecular form
(panel c). The increased H$_{2}$ cooling rate then roughly equals the
adiabatic heating rate, and thus the highest density gas is nearly
isothermal. As the gas is transformed into molecular form, the equation of
state changes as well so that $\gamma_{\rmn ad}$ evolves from 5/3 to
approximately 7/5 (panel d).
}
\label{nh_phase}
\end{figure*}

\begin{figure*}
\includegraphics[width=.8\textwidth]{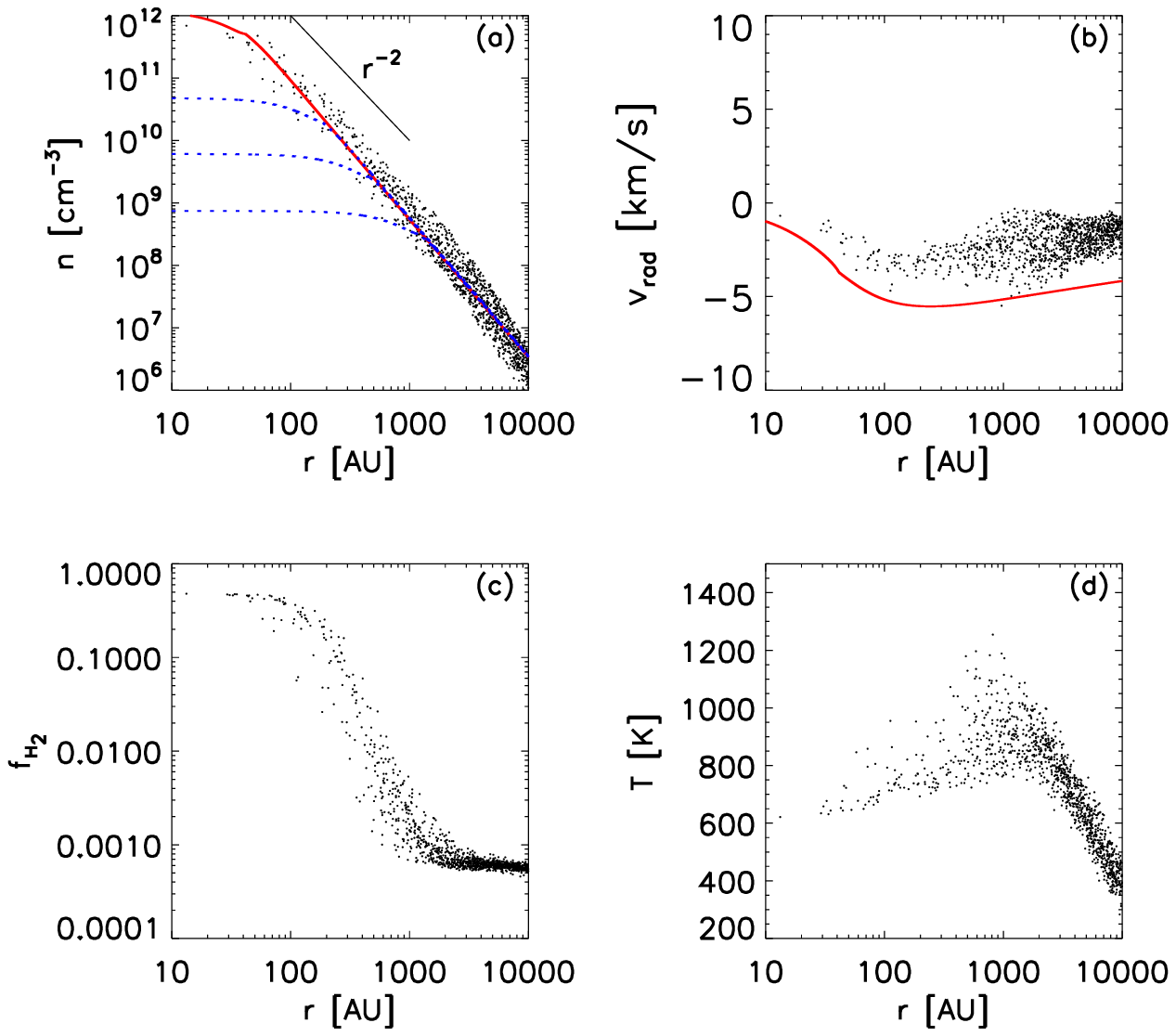}
\caption{Initial collapse prior to the formation of the first sink particle. Shown are a number of variables as a function of radial distance from the density maximum. {\it (a)} Number density; {\it (b)} radial velocity; {\it (c)} molecule fraction; and {\it (d)} temperature vs. $r$. Also shown are the number density and radial velocity of a Larson-Penston type similarity solution for the equation of state $P = K\rho^{\gamma}$, where $\gamma = 1.09$ and $K = 4.2 \times 10^{11}$\,(in cgs units), as in Omukai \& Nishi (1998). The solid red lines (in panels a and b) represent the latest stage of collapse that was calculated, when the peak density is near 10$^{12}$\,cm$^{-3}$, while the dashed blue lines (in panel a) represent various earlier stages of collapse. Note that the analytical solution reproduces the density profile found in the numerical calculation very well.  The velocity profile is also in reasonable agreement with the simulation, though, also as in Omukai \& Nishi (1998), exhibiting a somewhat larger normalization due to differences in initial and boundary conditions.}
\label{rad_prof}
\end{figure*}

\section{Physical Ingredients}

\subsection{Protostellar accretion}
Unlike the gas from which present-day stars are formed, which can typically cool to $\sim$ 10 K,  primordial gas contained no metals and could only cool down to $\sim$ 100 K through H$_{2}$ and HD line transitions. These elevated temperatures lead to a higher accretion rate, which can be estimated on dimensional grounds by assuming that the Jeans mass is infalling at the free-fall rate, resulting in
\begin{equation}
\dot{M} \simeq \frac{c_{\rmn s}^{3}}{G} \propto T^{3/2}  \mbox{\ ,}
\end{equation}

\noindent where c$_{\rmn s}$ is the soundspeed.  This is quite close to the \cite{shu1977} similarity solution for collapse of a singular isothermal sphere, which gives $\dot{M} = 0.975 c_{\rmn s}^{3}/G$.  The Shu solution begins with an isothermal cloud with $\rho \propto r^{-2}$ and initially negligible infall velocities. The solution then describes the subsequent `inside-out' collapse of the cloud, which physically corresponds to accretion of mass onto the central protostar.  The earlier similarity solution of \cite{larson1969} and \cite{penston1969} described the collapse of an initially uniform cloud into a peaked structure with an $r^{-2}$ density profile. This solution was later expanded by \cite{hunter1977} to capture the evolution after the initial formation of a hydrostatic core when it grows via accretion of the envelope. At the point of initial protostar formation, corresponding to when the central density becomes infinite in the similarity solution, the infall velocity is already 3.3 times the sound speed, and the accretion rate grows to 48 times higher than that found in Shu collapse (\citealt{hunter1977}).  

In our simulation, the collapse of the gas from initially more uniform densities can be expected to behave similarly to a Larson-Penston-type solution (e.g. \citealt{omukai&nishi1998}).  The following accretion of this gas onto the central hydrostatic core should proceed from the `inside-out' at rates similar to those found in the Shu and Larson-Penston-Hunter solutions. 

\subsection{Protostellar feedback effects}  
As the protostellar mass grows through accretion, feedback from photoionization, heating, and radiation pressure will begin to have important effects on the accretion flow and may set the upper mass limit for both present-day and primordial stars.  One important difference between feedback in primordial and present-day massive star formation is due to the existence of dust grains in the latter case.  For instance, \cite{ripamontietal2002} found with their one-dimensional calculations that Pop III protostellar radiation pressure will not affect the initial collapse and early stages of accretion due to the low opacity of the infalling gas.  In contrast, for present-day star formation the increased opacity from dust enhances the radiation pressure and can halt infall onto stars of more than a few tens of solar masses without further mechanisms to alleviate the radiative force (see summary in \citealt{mckee&ostriker2007}). On the other hand, later in the life of a present-day star, dust absorption of ionizing photons will alter the evolution of the H\,II region around the star, thus reducing the photoionization feedback.    

\cite{omukai&palla2003} performed a detailed study of spherical accretion of metal-free gas onto a growing hydrostatic core.  In this case, electron scattering is the major source of opacity determining the Eddington luminosity $L_{\rmn EDD}$ and the critical accretion rate $\dot{M}_{\rmn crit}$, above which the protostellar luminosity exceeds $L_{\rmn EDD}$ before the protostar reaches the zero-age main sequence (ZAMS), disrupting further accretion.  As estimated in \cite{bromm&loeb2004}, if we assume the protostellar luminosity before the onset of hydrogen burning is approximately the accretion luminosity, $L_{\rmn acc} \simeq GM_{*}\dot{M}/R_{*}$, and set this luminosity equal to $L_{\rmn EDD}$, we find that    

\begin{equation}
\dot{ M}_{\rmn crit} \simeq \frac{L_{\rmn EDD} R_*}{G M_*} \sim 5 \times 10^{-3} \, \rmn {M_{\odot} yr^{-1}}  \mbox{\ ,}
\end{equation}  

\noindent where $R_*$ $\sim$ 5\,R$_{\odot}$ (\citealt{bromm&kudritzki&loeb2001}).  This approximation is quite close to that found in the calculations of \cite{omukai&palla2003}.  It should be noted that initial accretion rates higher than $\dot{M}_{\rmn crit}$ are possible since protostars begin with much larger radii and only gradually shrink during Kelvin-Helmholtz contraction.  

Other types of feedback also have the potential to halt protostellar accretion and set the upper mass limit of Pop III stars.  \cite{mckee&tan2008} consider several feedback mechanisms operating during Pop III star formation, including H$_2$ photodissociation and  Ly$\alpha$ radiation pressure.  They find, however, that the accretion rate is not significantly reduced through feedback until HII region breakout beyond the gravitational escape radius, generally occuring once the star has reached 50-100 M$_{\odot}$.  \cite{omukai&inutsuka2002}, on the other hand, find that for spherically symmetric accretion onto an ionizing star, the formation of an HII region in fact does not impose an upper mass limit to Pop III stars.  As \cite{mckee&tan2008} determine, however, this does not remain the case for rotating inflow.      

This previous work shows that further considerations such as a time-dependent accretion rate and a non-spherical, disk-like geometry around the protostar (see also \citealt{tan&mckee2004}) will be integral to fully determining the effects of protostellar feedback and will require study through three-dimensional calculations. We do not include any radiative feedback effects here, and thus our work represents a no-feedback limit of protostellar accretion onto a Pop III star.  However, during the initial stages of protostellar growth, the feedback should be much weaker than in later stages when the star becomes more massive and luminous.  As mentioned above, \cite{mckee&tan2008} estimate that the accretion rate is not significantly reduced until the Pop III star has gained $\sim$ 50 M$_{\odot}$, and the rate of protostellar mass growth given by \cite{bromm&loeb2004} indicates that such masses are not reached until $\sim$ 10$^4$ yr after accretion begins.  We thus end our calculation somewhat before this, after 5000 yr of accretion, before feedback effects have become too strong to ignore.  Up until this point, our calculations should still give a good estimate of the geometry and possible fragmentation of the prestellar core as well as the early protostellar accretion rate and its time dependence.

\subsection{Chemistry, heating and cooling}

The chemistry, heating and cooling of the primordial gas is treated in a fashion very similar to previous studies (e.g. \citealt{bromm&loeb2004,yoshidaetal2006}).  We follow the abundance evolution of H, H$^+$, H$^-$, H$_2$, H$_{2}^{+}$, He, He$^+$, He$^{++}$, e$^-$, and the deuterium species D, D$^+$, D$^-$, HD, and HD$^+$.  We use the same chemical network as used in \citet{johnson&bromm2006} and include the same cooling terms.   

At densities greater than $\simeq 10^8$ cm$^{-3}$, three-body processes gain importance.  The reaction rates for

\begin{equation}
\rmn{H + H + H \rightarrow H_2 + H}
\end{equation}

\noindent and

\begin{equation}
\rmn{H + H + H_2 \rightarrow H_2 + H_2}
\end{equation}

\noindent must be included (\citealt{pallaetal1983}). Because the gas is becoming fully molecular, in our cooling rates we furthermore account for enhanced cooling due to collisions between H$_2$ molecules.  The heating rate due to H$_2$ formation heating is also included.  Furthermore, the equation of state of the gas must be modified.  More specifically, we adapt the adiabatic exponent $\gamma_{\rmn ad}$ and the mean molecular weight $\mu$ to account for the increasing H$_2$ abundance $f_{\rmn H_{2}}$.  These modifications for densities greater than $10^8$ cm$^{-3}$ are applied in the same way as described in \citet{bromm&loeb2004}, except for the modification of $\gamma_{\rmn ad}$, which is handled in the same way as decribed in \citet{yoshidaetal2006}.     

\begin{figure*}
\includegraphics[width=.4\textwidth]{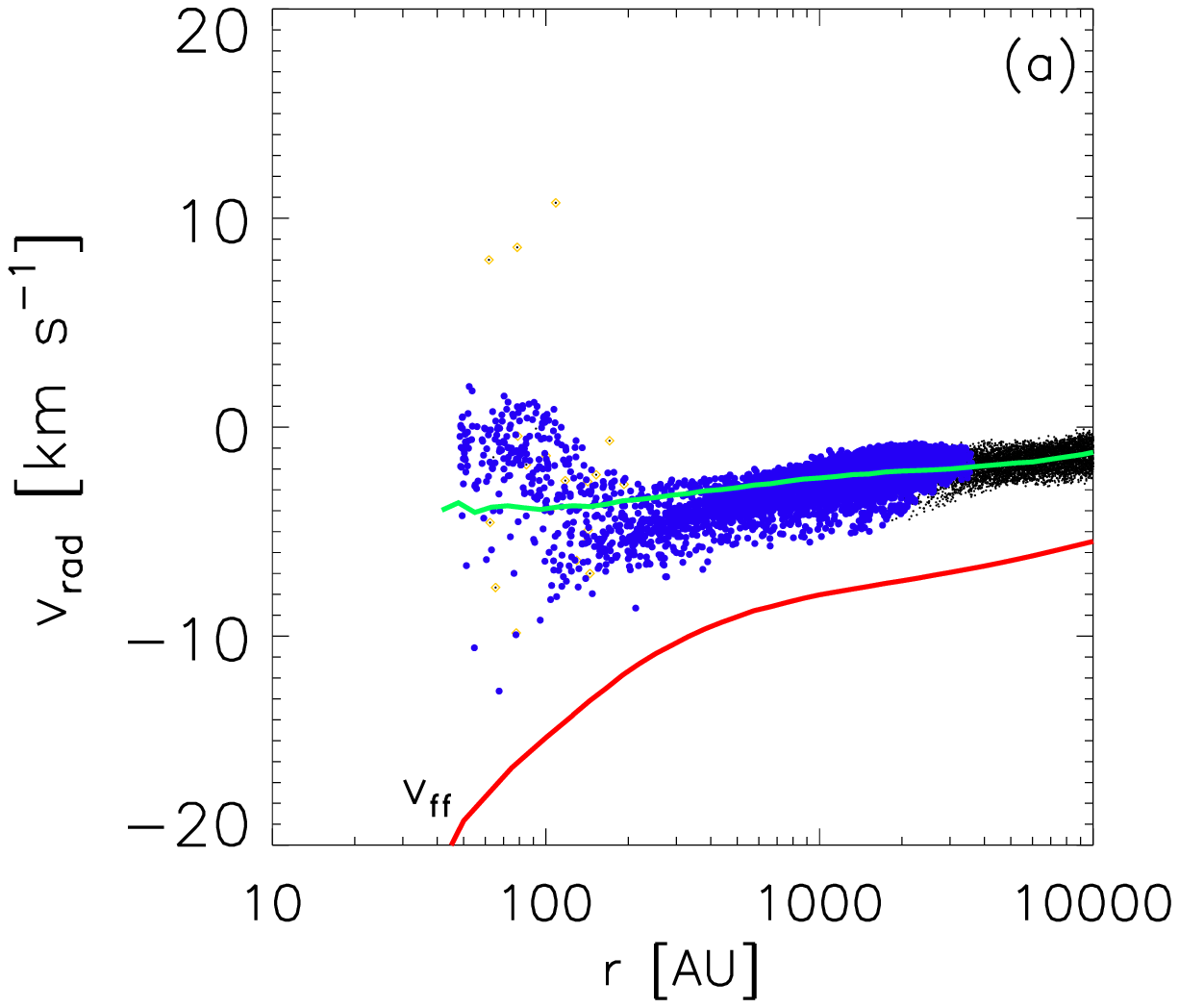}
\includegraphics[width=.4\textwidth]{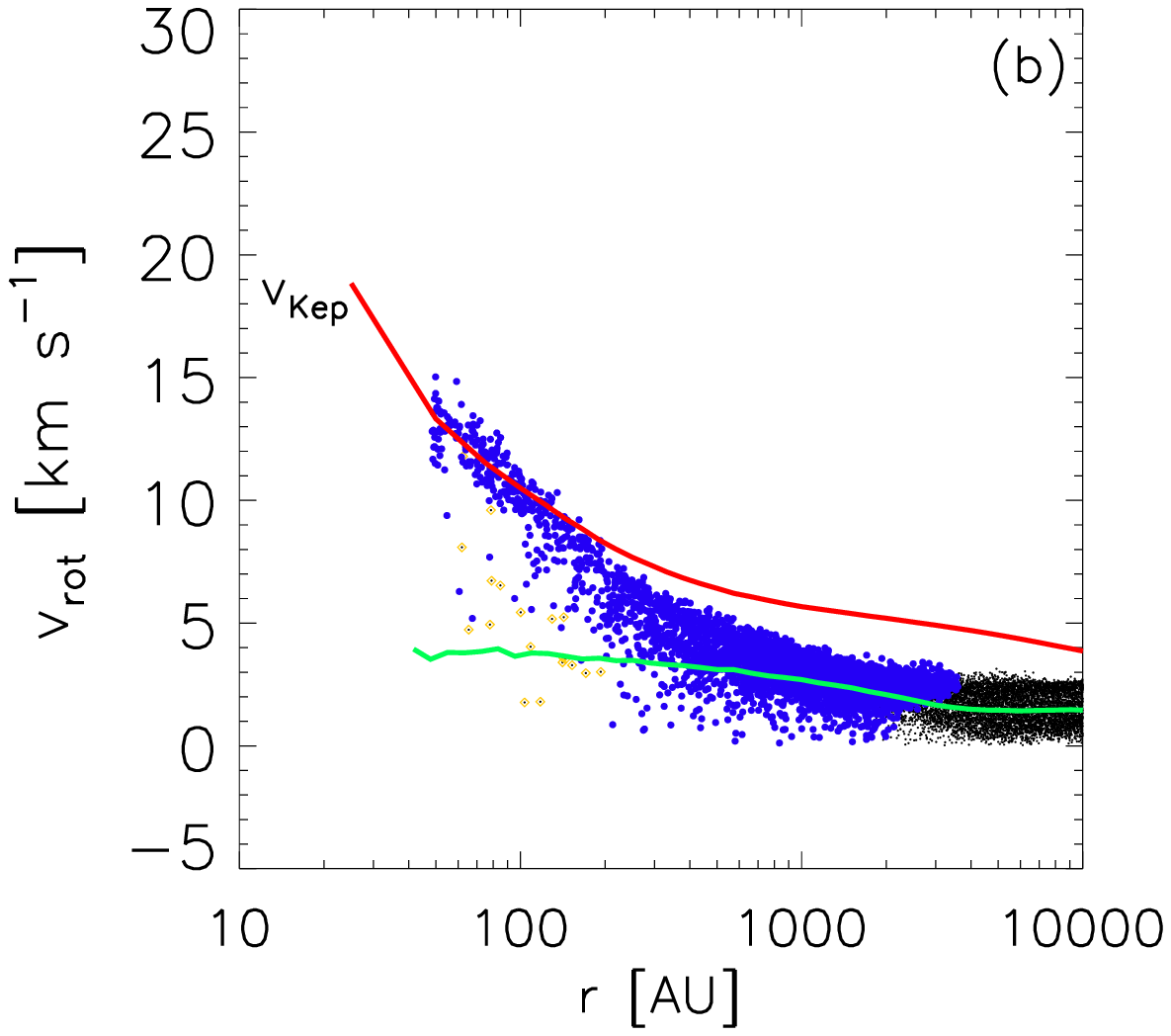}
\includegraphics[width=.4\textwidth]{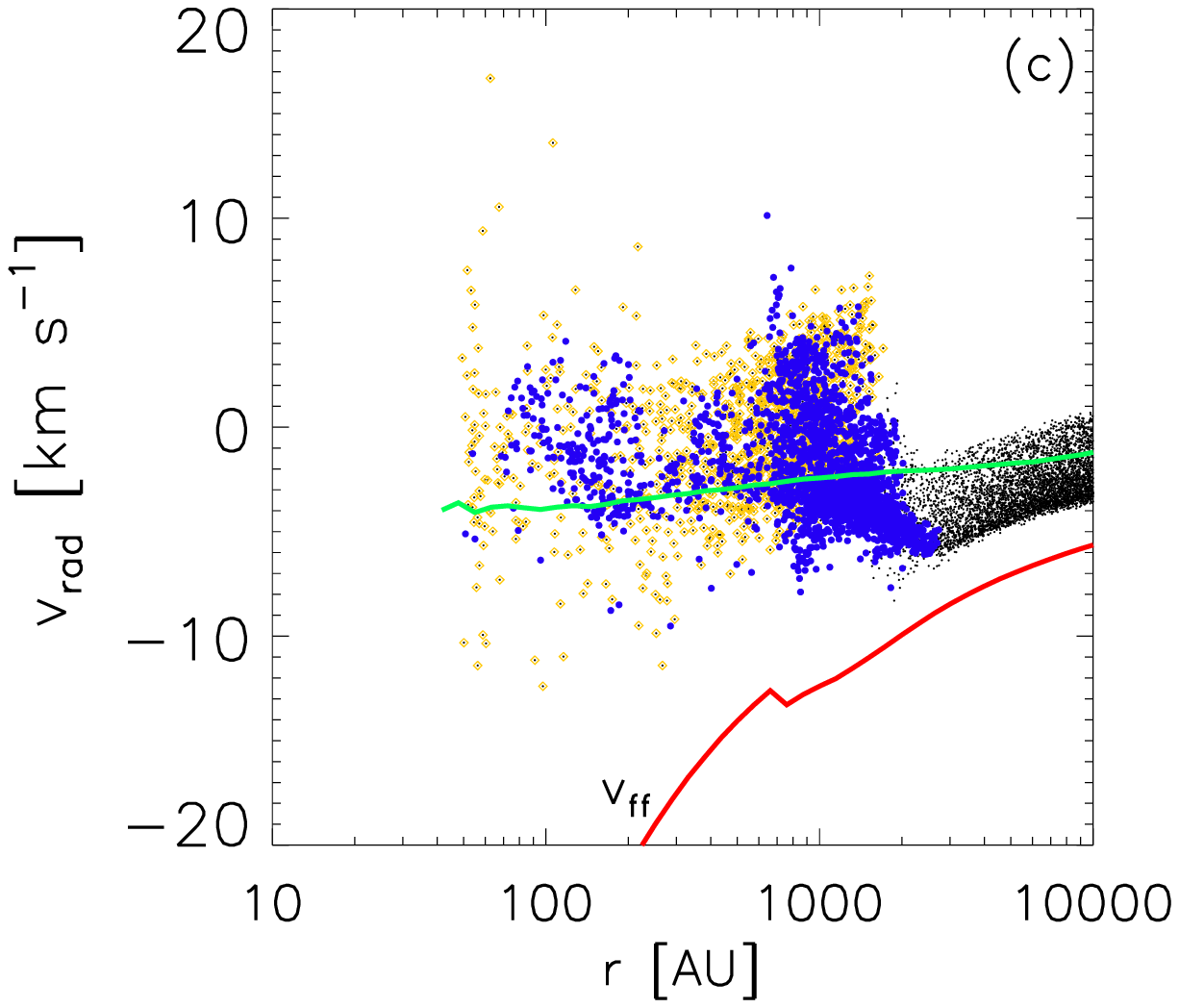}
\includegraphics[width=.4\textwidth]{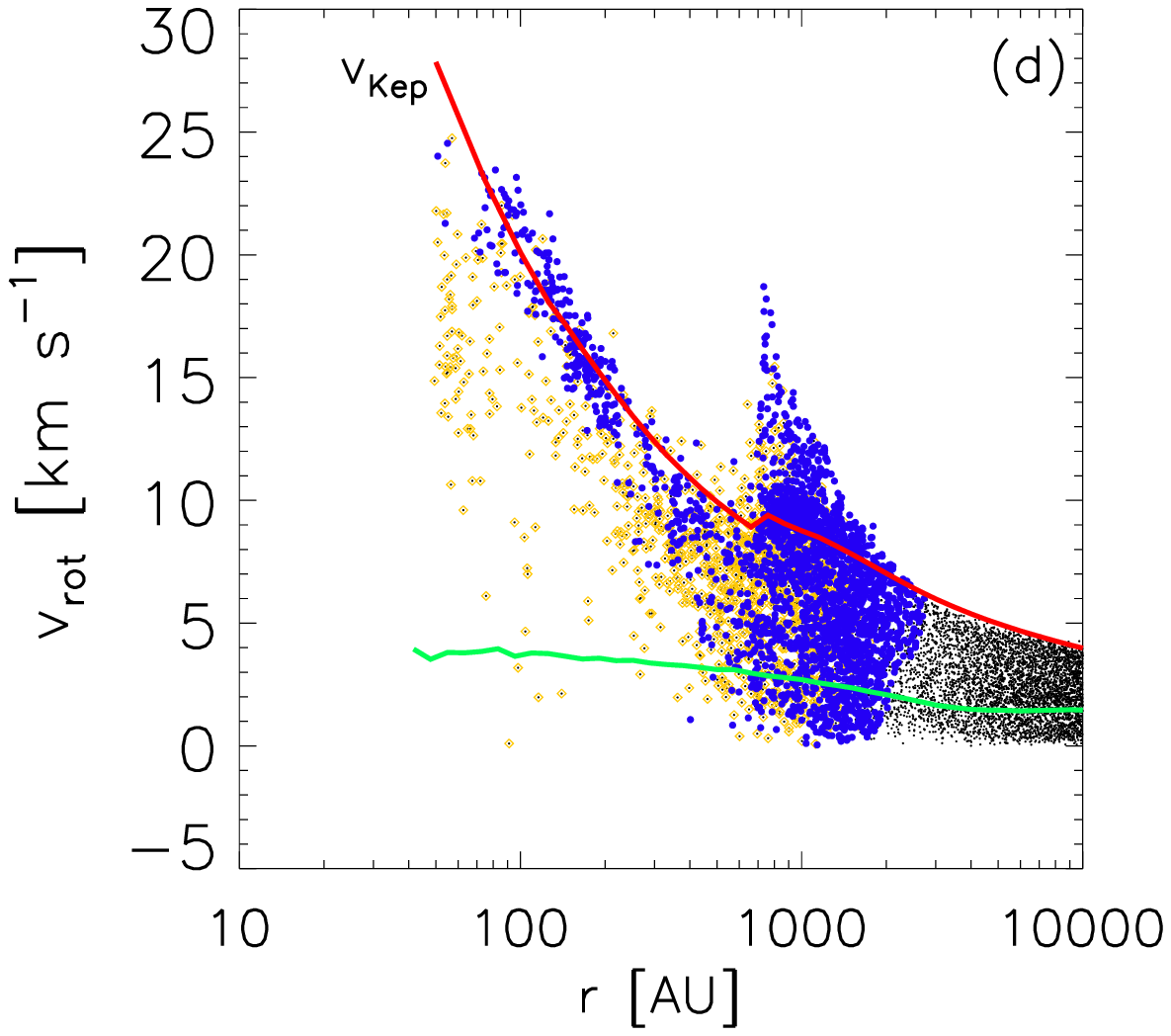}
\caption{Kinematics around the main sink particle. {\it(a)}  Radial velocity versus distance from the main sink at 250 yr after the initial sink formation.  Colored particles are those with  densities greater than 10$^8$ cm$^{-3}$ - yellow diamonds are high-density heated particles with temperatures greater than 2500 K, and blue circles are high-density cool particles with temperatures less than 2500 K. The red line shows the gravitational free-fall velocity $v_{\rmn ff}$, determined using the enclosed mass within the given radius.  For comparison to the physical situation at earlier times, the green line is the radially averaged radial velocity around the main sink immediately after it first forms.  {\it(b)} Rotational velocity versus distance from the main sink at 250 yr after sink formation. Blue and yellow particles have the same meaning as in panel~(a).  The red line shows $v_{\rmn Kep}$, calculated using the enclosed mass. The green line is the radially averaged rotational velocity around the main sink immediately after it first forms.  {\it(c)} Radial velocity versus distance from main sink at 5000 yr after initial sink formation.  Notation is the same as in panel {\it(a)}.  {\it(d)} Rotational velocity versus distance from main sink at 5000 yr after initial sink formation. We adopt the same convention as in panel~(b). Towards later times, the cold gas settles into a nearly Keplerian configuration.}
\label{vel_part}
\end{figure*}

\section{Numerical Methodology}
\subsection{Initial setup}
We carry out our investigation using Gadget, a widely-tested three dimensional smoothed particle hydrodynamics (SPH) code (\citealt{springeletal2001,springel&hernquist2002}). Simulations are performed in a periodic box with size of 100~$h^{-1}$~kpc (comoving) and initialized at $z=99$ with both DM particles and SPH gas particles.  This is done in accordance with a $\Lambda$CDM cosmology with $\Omega_{\Lambda}=0.7$, $\Omega_{\rmn M}=0.3$, $\Omega_{\rmn B}=0.04$, and $h=0.7$.  
We use an artificially large normalization of the power spectrum,
$\sigma_8=1.4$, to accelerate structure formation in our relatively small
box.
 As discussed in Section 5, our high value of $\sigma_8$ is expected to have little effect on our overall qualitative results, though this will be explored further in future work.

A preliminary run with a modest resolution of $N_{\rmn SPH}=N_{\rmn DM}=128^3$ allows us to locate the formation site of the first minihalo.
We then employ a standard hierarchical zoom-in procedure to achieve the necessary high-mass resolution in this region (e.g., \citealt{navarro&white1994,tormenetal1997,gaoetal2005}).  This method was also used in the work of \cite{greifetal2008}, in which they simulated the formation of a first galaxy beginning from cosmological initial conditions.  Applying this method for our calculation involved beginning the simulation again at $z=99$, but now with an additional three nested refinement levels of length 40, 30, and 20 kpc (comoving) centered on the site where the first minihalo will form.  This high-resolution run is thus initialized at high redshift with the same initial fluctuations as in the preliminary low-resolution run, though the high-resolution region further includes fluctuations at smaller wavelength scales as allowed by the increased Nyquist frequency.

Particle positions are shifted so that the high resolution region will be in the center of the simulation box. Each level of higher refinement replaces particles from the lower level with eight child particles such that in the final simulation a parent particle is replaced by up to 512 child particles. The highest-resolution gas particles each have a mass $m_{\rmn SPH} =0.015$~M$_{\odot}$, so that the mass resolution of the refined simulation is: $M_{\rmn res}\simeq 1.5 N_{\rmn neigh} m_{\rmn SPH}\la  1$ M$_{\odot}$, where $N_{\rmn neigh}\simeq 32$ is the typical number of particles in the SPH smoothing kernel (e.g. \citealt{bate&burkert1997}).  

The size of the refinement levels had to be chosen carefully.  The level of highest refinement must be large enough to encompass all the mass that will eventually become the first minihalo.  If this level of highest refinement is too small, lower-resolution particles will drift into the center of the collapsing region and disrupt the collapse of the high-resolution particles.  Furthermore, there will be some artificial density enhancement for lower-resolution particles that are near the boundary of a higher-resolution region.  This is because the particles of lower refinement levels have higher masses, and for a given density the corresponding smoothing length for lower-resolution particles should be larger than the smoothing length for higher-resolution particles.  Smoothing lengths for particles of both resolutions will be very similar, however, if they are very close to each other.  This causes low-resolution particles near a boundary to have artificially reduced smoothing lengths and overly enhanced densities.  However, this will not pose a problem as long as the highest refinement level is large enough such that the low-resolution particles are well outside of the collapsing region of interest.      

\subsection{Sink particle creation}
When the density of a gas particle becomes greater than $n_{\rmn max}=10^{12}$\,cm$^{-3}$, this particle and the surrounding gas particles within a distance of less than the accretion radius $r_{\rmn acc}$ are converted into a single sink particle. 
We set $r_{\rmn acc}$ equal to the resolution length of the simulation, $r_{\rmn acc} =  L_{\rmn res} \simeq 50$\,AU, where:
\begin{displaymath}
L_{\rmn res}\simeq 0.5 \left(\frac{M_{\rmn res}}{\rho_{\rmn max}}\right)^{1/3} \mbox{\ ,}
\end{displaymath}
with $\rho_{\rmn max}\simeq n_{\rmn max}m_{\rmn H}$.
The sink particle's initial mass is close to the simulation's resolution mass, $M_{\rmn res} \simeq 0.7$ M$_{\odot}$. 

Our density criterion for sink particle formation is robust even though we do not explicitly check that the sink region satisfied further criteria such as being gravitationally bound and not having rotational support.  As discussed in \cite{brommetal2002}, crossing the $n_{\rmn max}$ threshold is already equivalent to meeting these further criteria due to the runaway nature of the collapse.  The average density of the disk (see Sec. 4.2.1) does not go above $\simeq$ 10$^{10}$ cm$^{-3}$, and gas must collapse to densities two orders of magnitude higher before a sink is formed at the center of the infalling region.  This along with our very small value for $r_{\rmn acc}$ ensures that the particles that become sinks are indeed  undergoing gravitational collapse.  Furthermore, an examination of movies of the simulation shows that even the lowest-mass sinks are created only in pre-existing high-density clumps of gas that formed through gravitational instability, even though the clumps may no longer be clearly visible after the sink forms and accretes mass.  
During earlier test runs, however, using only the density criterion was problematic for other numerical reasons.  This was because the larger sinks would contribute their high masses to the kernels of nearby gas particles, causing them to artificially jump to $n>10^{12}$\,cm$^{-3}$ and become sink particles themselves.  These artificial sink particles had masses well below $M_{\rmn res}$.  However, once we corrected this problem by removing all sink particles from the density calculation, all sinks that were formed had initial masses $\ge$ $M_{\rmn res}$.    

After a sink's formation, its density, temperature, and pressure are no longer evolved in time.
Instead, their density is held constant at 10$^{12}$ cm$^{-3}$.  The temperature of the sinks is also held constant at 650 K, a typical temperature for gas reaching 10$^{12}$ cm$^{-3}$, and the sinks are furthermore kept at the pressure which corresponds to the above temperature and density.  Giving the sink a temperature and pressure makes it behave more like a gas particle, and it furthermore prevents the existence of a pressure vacuum around the sink that would make the sink accretion rate artificially high (see \citealt{brommetal2002,marteletal2006}).  Changes in the sink's position, velocity, and acceleration due to gravitational interaction are still evolved, however, and as the sink accretes mass its gravity becomes the dominant force, and it gradually begins to act more like a non-gaseous N-body particle.  

Once a sink particle is formed, all gas particles or other sink particles that cross within  $r_{\rmn acc}$ of the sink particle are removed from the simulation, and their mass is added to that of the sink particle.  After each subsequent accretion event, as well as when the sink is initially formed, the position and velocity of the sink are set to be the mass-weighted average of the sink particle and the gas particles that it has accreted.  The sink approximately represents the growing hydrostatic core, though its size, of order $r_{\rmn acc}$, is much larger than the true expected size of a core (e.g. \citealt{omukai&nishi1998}).

The sink particle method has multiple advantages.  It eliminates the need to include high density physics that becomes relevant only at  $n>10^{12}$ cm$^{-3}$.  Furthermore, simulation timesteps become prohibitively small as the density increases, and representing a region of  $n>10^{12}$ cm$^{-3}$ with a single sink particle allows the mass flow into protostellar regions to be followed for many dynamical times without continuing to increasingly high densities and small timesteps.  Instead of estimating the accretion rate from the instantaneous density and velocity profiles at the end of the simulation, as done in \cite{abeletal2002} and \cite{yoshidaetal2006}, the sink particle method allows the accretion history to be directly followed.  Furthermore, without the sink particle method, finding multiple fragments would require that they form at very nearly the same time, such as the fragments in \cite{turketal2009} which formed only 60 years apart.  Using sink particles allows for the study of fragmentation that occurs on significantly longer timescales than this.            

\begin{figure*}
\includegraphics[width=.9\textwidth]{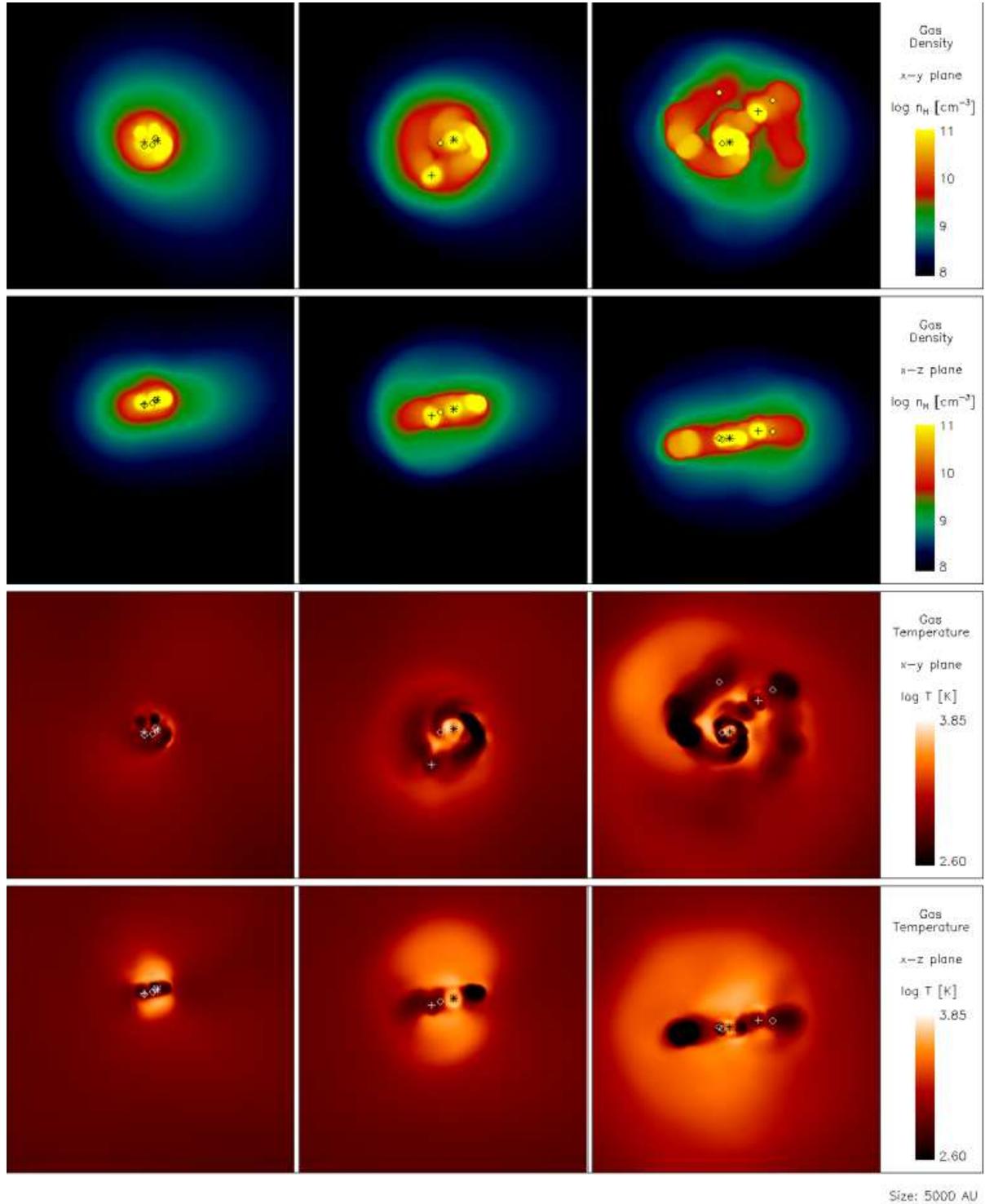}
\caption{Density and temperature projections of the central 5000 AU.  Each row shows the projections at 1000 yr (left), 2000 yr (center), and 5000 yr (right) after the
initial sink formation. Asterisks denote the location of the most massive sink.  Crosses show the location of the second most massive sink.  Diamonds are the locations
of the other sinks.  $\itl{Top}$ $\itl{ row:}$ Density structure of the central region in the x-y plane. $\itl{Second}$ $\itl{row:}$ Density structure of the central
region in the x-z plane.  $\itl{Third}$ $\itl{row:}$ Temperature structure of the central region in the x-y plane.  $\itl{Bottom}$ $\itl{row:}$ Temperature structure
of the central region in the x-z plane.}
\label{morph}
\end{figure*}

\section{Results}
\subsection{Initial runaway collapse}

Fig.~\ref{zoom_in} shows the final simulation output on various scales.  The expected filamentary structure is apparent, and the first minihalo has formed in the central region where several filaments intersect by $z \simeq 20$.  We measure the halo virialization radius $r_{\rmn vir}$ as the point at which $\rho_{\rmn DM} =  \rho_{\rmn vir} \equiv 200\rho_{\rmn b}$, where $\rho_{\rmn DM}$ is the average halo DM density, $\rho_{\rmn vir}$ is the DM density at the point of virialization, and $\rho_{\rmn b}$ is the average background matter density at the time of halo virialization.  We then find that  $r_{\rmn vir}$ $\simeq$ 90 pc and $M_{\rmn halo}~\simeq$~2.6~$\times$~10$^5$~M$_{\odot}$.  The spin parameter $\lambda = J |E|^{1/2}/{\rmn G}M^{5/2}$ is approximately 0.13 for our halo, where $J$ is the total angular momentum, $E$ is the binding energy, and $M$ is the total halo mass.  For our minihalo $\lambda$ is higher than the typical value of 0.05 (e.g. \citealt{barnes&efstathiou1987,jh2001}).  
 This large spin may result from our enhanced value of $\sigma_8$.  However, as discussed in Section 5, the angular momentum profile of the central gas in our simulation is still quite similar to that found from previous cosmological simulations initialized with a lower $\sigma_8$.  Thus, our overall results concerning disk evolution and fragmentation are not expected to significantly change with a lower value of minihalo spin or  $\sigma_8$.

After the DM in the minihalo virializes, the gas continues to collapse.  As shown in Fig.~\ref{nh_phase}, the results of our calculation up to the point of creation of the first sink particle agree well with previous studies (e.g. \citealt{bromm&larson2004,yoshidaetal2006}).  As expected, the gas heats adiabatically until reaching  $n\sim 1$ cm$^{-3}$, where it attains maximum temperatures of $\sim$ 1000 K.  After this point the gas begins to cool through H$_2$ rovibrational transitions, reaching a minimum temperature of  $T\sim 200$\,K at a density of $n\sim 10^4$\,cm$^{-3}$.  This is the density at which the gas reaches a quasi-hydrostatic `loitering phase' (\citealt{brommetal2002}).  After this phase the gas temperature rises again due to compressional heating until  $n\sim 10^8$ cm$^{-3}$, when three-body processes become important and cause the gas to turn fully molecular by  $n\sim 10^{12}$\,cm$^{-3}$ (see detailed discussion in \citealt{yoshidaetal2006,yoh2008}).  

Fig.~\ref{rad_prof} shows the radial structure of the gas surrounding the density peak just before the first sink particle is created (compare with figure~2 in \citealt{bromm&loeb2004}). The gas inside the central 100\,AU has become almost fully molecular, and it evolves roughly isothermally due to the approximate balance between compressional and H$_2$ formation heating and enhanced H$_2$ cooling. The deviation from spherical symmetry of this central region is apparent in the `thickness' of the radial profiles, particularly the profile of the H$_2$ fraction $f_{\rmn H_{2}}$.  However, the gas density and velocity profiles still have the same general properties as those of the spherically symmetric Larson-Penston type similarity solutions, an example of which is shown in Fig.~\ref{rad_prof}. The good agreement is
evident, at least for the initial collapse phase.
\cite{omukai&nishi1998} used the same solution to describe their one-dimensional radiation-hydrodynamics simulations of primordial protostar formation.  Though the velocities in the similarity solution are normalized to higher values, the velocites in our three-dimensional calculation exhibit very similar overall behavior, and the density profile matches the similarity solution especially well.

\subsection{Protostellar accretion}
\subsubsection{Disk formation}

We next discuss the transition from the quasi-spherical initial collapse
towards a disk-like configuration. This transition is clearly visible in
the velocity structure of the central region, which changes dramatically over the last several thousand years of the simulation as the disk-like nature of the central region becomes more apparent over time. In Fig.~\ref{vel_part}, we illustrate the evolution of the radial (left panels) and rotational velocity (right panels), measured with respect to the most massive sink. In comparing to the free-fall velocity, we calculate $v_{\rmn ff}$ using the enclosed mass $M_{\rmn enc}$ within a radial distance $r$ from the main sink. At both times shown, 250 and 5000~yr after initial sink formation, pressure support from the central gas leads to a deviation from pure free fall, but the overall behavior of the radial velocity of the outer particles is similar to that of the $v_{\rmn ff}$ profile. This is not the case for the particles very near the sink, however. The average radial velocity of the central gas particles, particularly the cold ones (shown in blue), approaches zero as the gas begins to exhibit more rotational motion.  

The central region is evolving into a Keplerian disk ($v_{\rmn rot}$~$\propto$~$r^{-1/2}$) by just 250 yr after the first sink has formed. Note the comparison between the particle rotational velocities $v_{\rmn rot}$ and the Keplerian velocity $v_{\rmn Kep}$ (red line in Fig.~\ref{vel_part}), defined as

\begin{equation}
 v_{\rmn Kep} = \sqrt{\frac{G  M_{\rmn enc}}{r}} \mbox{\ .}
\end{equation}

\noindent From Fig.~\ref{vel_part} (panel b), it is apparent that the inner 100-200 AU of the disk are becoming rotationally supported ($v_{\rmn rot}$~$\ga$~$v_{\rmn Kep}$) by 250\,yr after sink formation. Despite some amount of rotational support, however, a fraction of the particles is still able to approach the sink, thus allowing continued accretion.  
After 5000~yr, the disk has grown to a radial size of slightly less than 2000~AU (panel c), while the cold disk particles exhibit nearly Keplerian motion (panel d). The spike at 700 AU (panel d) corresponds to a group of particles rotating around the second largest sink. 

\begin{figure}
\includegraphics[width=.4\textwidth]{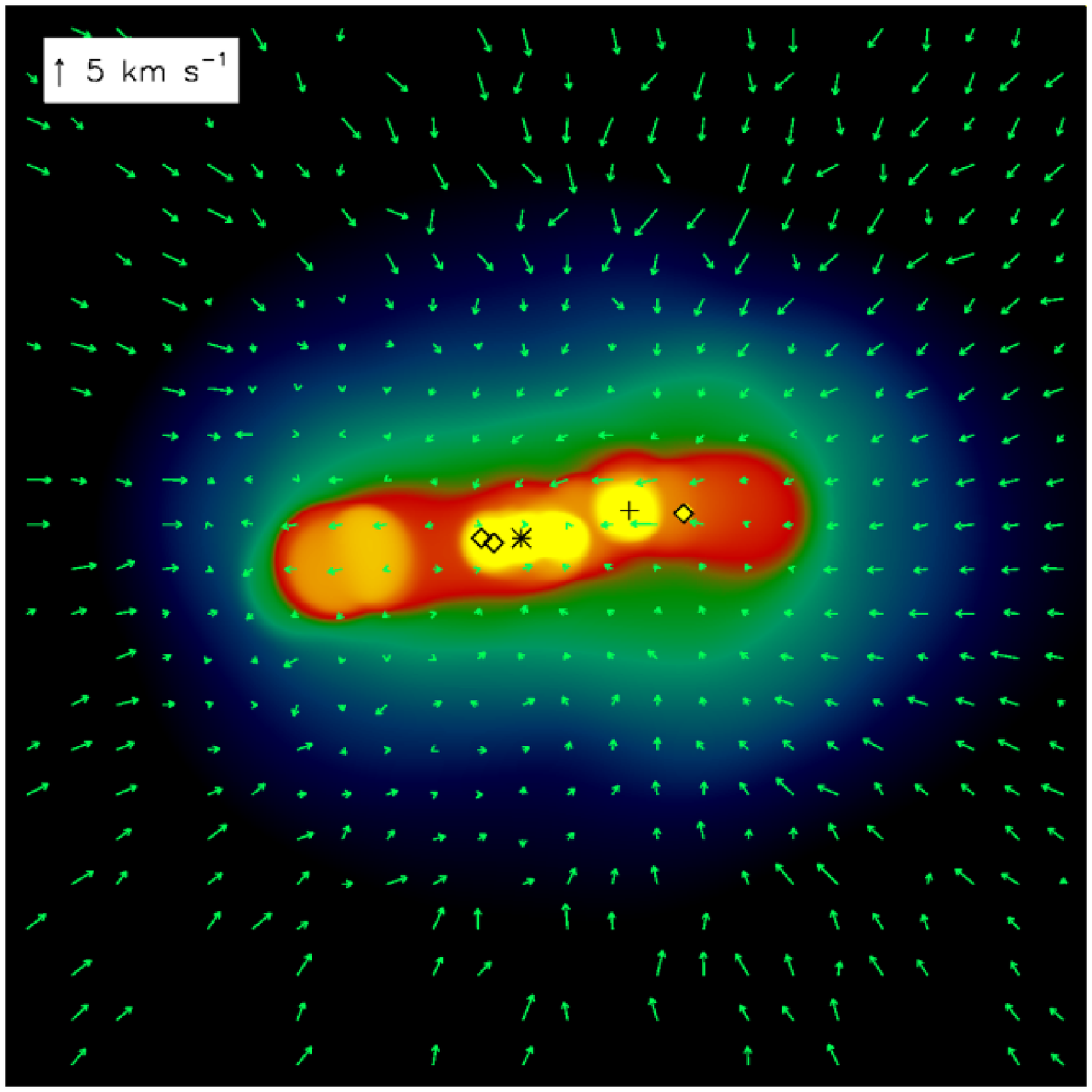}
\includegraphics[width=.4\textwidth]{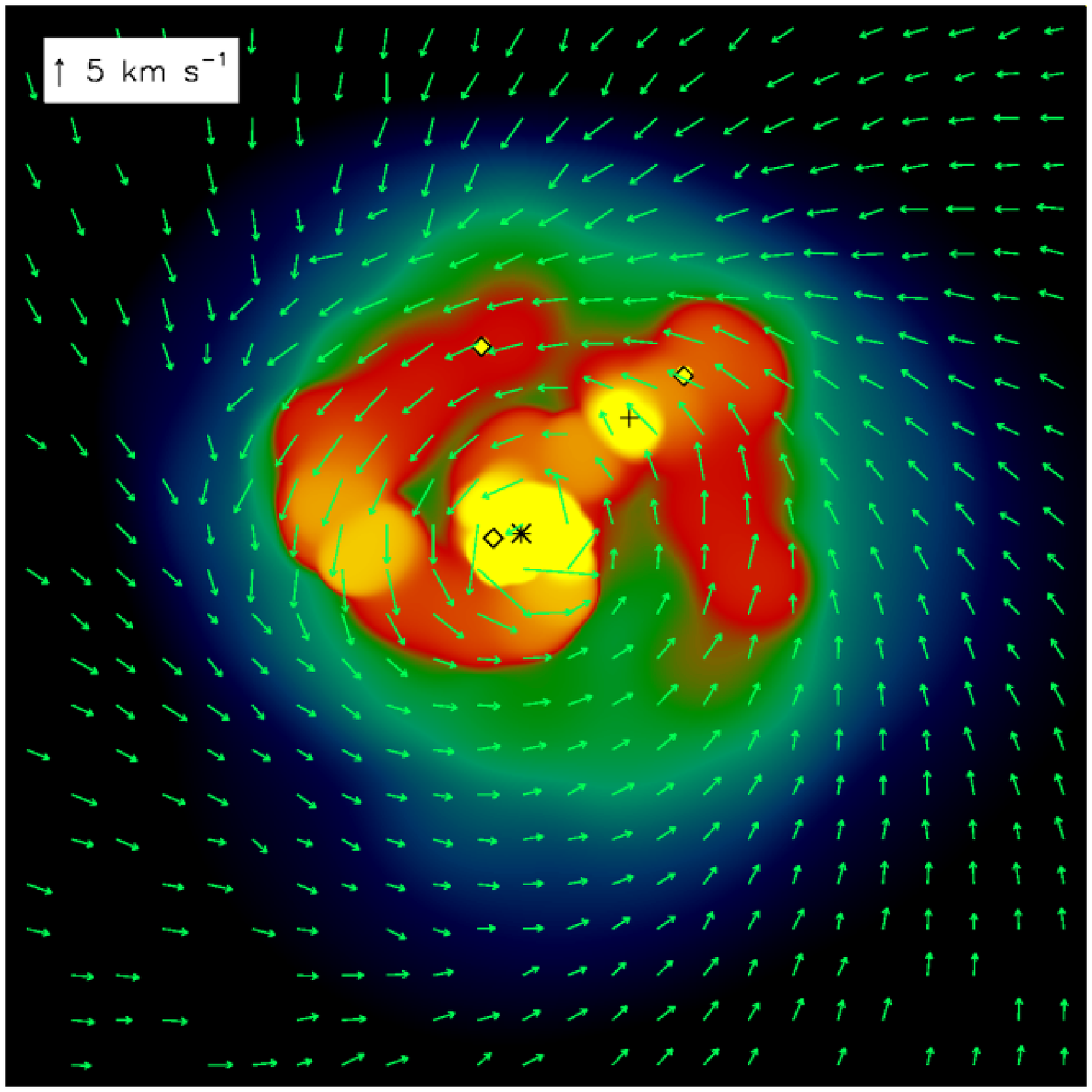}
\caption{Velocity field of the central gas distribution. Sinks are denoted as in Fig.~\ref{morph}. $\itl{Top:}$ Density projection in the x-z plane after 5000 yr, shown together with the velocity field.  Velocities are measured with respect to the center of mass of all  $n > 10^8$ cm$^{-3}$ particles.  $\itl{Bottom:}$ Same as what is shown in the top panel except in the orthogonal (x-y) plane.  It is evident how an ordered, nearly Keplerian velocity structure has been
established within the disk.}
\label{vel_arrows}
\end{figure}

The assembly of the disk-like configuration around the first sink particle,
and the subsequent fragmentation into additional sinks, are shown in
Fig.~\ref{morph}. The growing disk is unstable and develops a pronounced
spiral pattern, leading in turn to the formation of multiple high-density
peaks. As is evident, regions of high density correlate with those of low
temperature, due to the enhanced cooling there. The expanding, hour-glass
shaped, bubble of hot gas (clearly visible in the bottom row) reflects 
the release of gravitational energy when particles fall into the deepening
potential well around the central sink. The corresponding virial temperature
is
\begin{displaymath}
T_{\rmn vir}\simeq \frac{G M_{\rmn sink}m_{\rmn H}}{k_{\rmn B}r_{\rmn acc}}
\simeq 10^4 \mbox{\,K,}
\end{displaymath}
where $M_{\rmn sink}\sim 10 {\rmn \,M}_{\odot}$. This gravitational
source of heating causes the emergence of a hot gas phase around
the central sink ($T\simeq 7,000$~K), even in the absence of any direct
radiative feedback from the protostar which is not included here.
In Fig.~\ref{vel_arrows}, we illustrate the corresponding velocity field in two orthogonal planes at 5000~yr. The small infall velocities onto the disk and the high rotational, roughly Keplerian, velocites within the disk are apparent. 

\begin{figure}
\includegraphics[width=.45\textwidth]{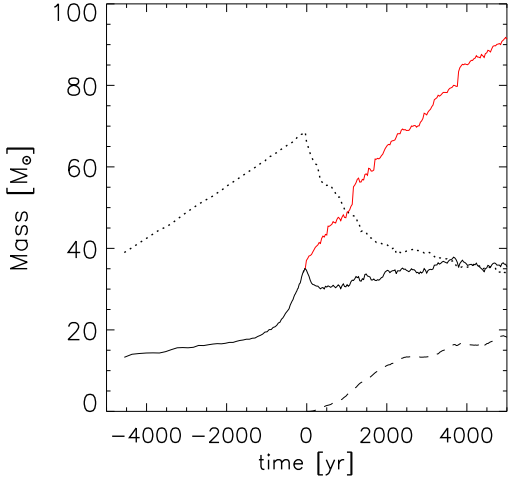}
\caption{Disk mass vs. time since the creation of the first sink particle (solid line).  Disk particles are considered to be all gas particles, excluding sinks, with density greater than 10$^8$ cm$^{-3}$ and a specific angular momentum within 50\% of $j_{\rmn Kep}$.  Once the first sink forms at $t = 0$~yr, the red line shows the total mass of the star-disk system.  Also shown is the total mass of $n >$ 10$^8$ cm$^{-3}$ particles (sinks excluded) with temperatures less than 2500 K (dotted line) and temperatures greater than 2500 K (dashed line).}
\label{diskmass}
\end{figure}

Fig.~\ref{diskmass} shows the growth of the disk over time, where $t=0$~yr marks the point at which the first sink forms. We estimate the disk mass by including all gas particles, sinks excluded, with a density greater than 10$^8$ cm$^{-3}$ and a specific angular momentum within 50\% of $j_{\rmn Kep}$, the specific angular momentum expected for a centrifugally supported disk, where
$j_{\rmn Kep} =   v_{\rmn Kep} r$. 
We here refer all velocities to the center of mass of the high-density ($n >$  10$^8$ cm$^{-3}$) particles.  In addition, we track the evolution of the mass in the `cool' (dotted line) and `hot' gas phase (dashed line), corresponding to temperatures less than 2500~K and greater than 2500~K, respectively. The mass of heated particles grows soon after the first sink is in place, whereas the cold component decreases as its mass is incorporated into the sinks or becomes part of the hot phase.  Eventually what remains of the cold component is confined to the disk itself (see bottom two rows of Fig.~\ref{morph}).   

The characteristics of the star-disk system in our simulation can be compared to the analytic predictions of \citet{tan&mckee2004}. At a density and temperature of 10$^8$ cm$^{-3}$ and 1000\,K, respectively, we find a value of $K'$ $\sim$ 1.33 from their equation (7), where  $K'$ is the entropy parameter, defined by

\begin{equation}
K' = \frac{P/\rho^{\gamma}}{1.88 \times 10^{12} \rmn cgs} = \left(\frac{T}{300\, \rmn K}\right)\left( \frac{n_{\rmn H}}{10^4 \rmn cm^{-3}}\right)^{-0.1} \mbox{\ .}  
\end{equation}

\noindent We estimate that the total star-disk system mass (red line in  Fig.~\ref{diskmass}) is $\sim$ 100 M$_{\odot}$ at the end of our simulation and set $\epsilon_{*d}$ equal to 1 for the no feedback case, where $\epsilon_{*d}$ is the fraction of infalling mass that is able to reach the star-disk system without being diverted by protostellar outflows, again in the terminology of Tan \& McKee.  From their equation (8), we then calculate an expected accretion rate onto the star-disk system of $\la$ 10$^{-2}$ M$_{\odot}$ yr$^{-1}$. In our simulation we find that over the 5000~yr of sink accretion, mass flows onto the star-disk system very steadily at approximately the same rate of $\sim$ 10$^{-2}$ M$_{\odot}$ yr$^{-1}$, which is also near the value for the average accretion rate onto the main sink particle (see Sec. 4.2.4). Fig.~\ref{diskmass} further shows that once the first sink forms, the disk mass stays fairly constant at $\sim$ 35 M$_{\odot}$ as mass falling from the outer regions onto the disk replaces the material accreted onto the sinks at similar rates.

\citet{tan&mckee2004} furthermore predict the time needed to build up a given stellar mass in their equation (9).  If we assume a value of 1 for $f_{\rmn d}$, the ratio of disk to stellar mass, then the time to build up a 43 M$_{\odot}$ star (the final mass of the main sink - see Sec. 4.2.4) is roughly 9000~yr.  The simulation yields a similar though smaller value of 5000 yr, which reflects the slightly higher accretion rate of our simulation.  Finally, if we estimate a value of 1 for $f_{\rmn Kep}$, the ratio of rotational to Keplerian velocity at the sonic point, then from equation (15) of \citet{tan&mckee2004} we find a predicted disk radius of $\sim$ 3000 AU.  This value is also similar to the value of $\sim$ 2000 AU found from the simulation.      

\begin{figure}
\includegraphics[width=.47\textwidth]{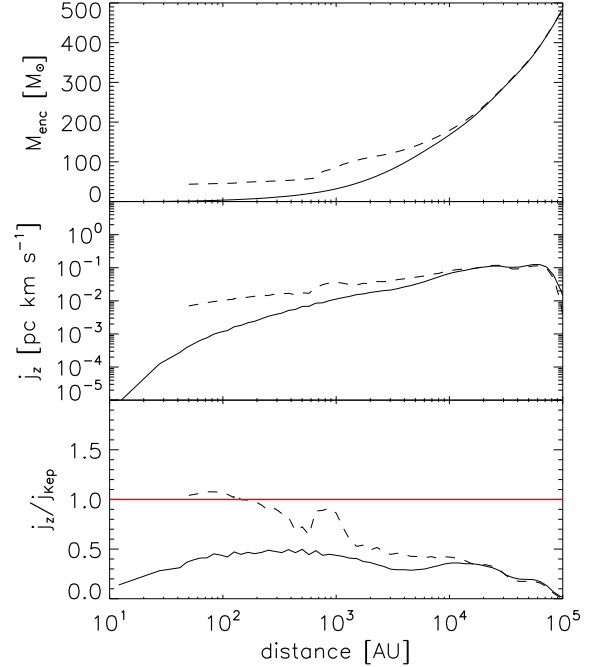}
\caption{Angular momentum structure. $\itl{Top:}$ Enclosed mass versus distance from main sink particle at last simulation output before sink formation (solid line) and after 5000 yr of sink accretion (dashed line).  $\itl{Middle:}$ Specific angular momentum $j_{z}$ of the cool (T $<$ 2500 K) gas with respect to distance from the main sink  at the same corresponding times.  $\itl{Bottom:}$ Degree of rotational support of the cool gas. The solid red line shows where $j/j_{\rmn Kep}$ = 1, the requirement for complete rotational support.}
\label{angmom}
\end{figure}

\subsubsection{Angular momentum evolution}
How does the distribution of angular momentum change during the simulation?
In Fig.~\ref{angmom}, we demonstrate the increase in rotational support of the central region over time. Shown are the evolution of the enclosed mass (top panel), the specific angular momentum along the z-axis, $j_{z}$ (middle panel), and the degree of rotational support (bottom panel). Here, rotational support is defined as

\begin{equation}
f_{\rmn rot} = \frac{j_{z}}{j_{\rmn Kep}}  \mbox{\ .}
\end{equation}   

\noindent We determine $j_{z}$ by averaging over all cool ($T < 2500$\,K) particles within small radial bins covering a total distance of 10$^5$\,AU.  Note how the inner 200\,M$_{\odot}$, corresponding to the central 10$^4$\,AU, have spun up significantly in the process of disk formation. This occurs as the inner mass shells fall towards the center, leading to the increase in rotational support.  The inner 200\,AU of cool gas are fully rotationally supported by the end of the simulation (bottom panel).  

To confirm that numerical viscosity is not a dominant factor in determining our results, we have compared the torque due to numerical viscosity ($\vec\tau_{\rmn visc}$) to those exerted by gravity and pressure ($\vec\tau_{\rmn grav}$ and $\vec\tau_{\rmn pres}$).  The total torque on a given particle within a gas cloud is given by

\begin{eqnarray}
\vec\tau_{\rmn  tot}&&=  \vec\tau_{\rmn grav} + \vec\tau_{\rmn pres}  + \vec\tau_{\rmn visc} \nonumber\\
&&= m_{\rmn part}\vec{r}_{\rmn part}\times(\vec{a}_{\rmn grav} + \vec{a}_{\rmn pres} + \vec{a}_{\rmn visc}) \mbox{\ ,}
\end{eqnarray}

\noindent where $\vec{r}_{\rmn part}$ is the distance of the particle from the center of the cloud, $m_{\rmn part}$ the particle mass, and $\vec{a}_{\rmn grav}$, $\vec{a}_{\rmn pres}$ and $\vec{a}_{\rmn visc}$ are the accelerations due to gravity, pressure and viscosity, respectively. Similar to \cite{yoshidaetal2006}, we store these accelerations for each particle and every timestep.  We find that torques from gravity and pressure are dominant by an order of magnitude except in regions near the sinks.  The viscous torques within $\sim$ 100 AU of the sinks may exceed the gravitational and pressure torques at any given instant.  However,  unlike $\vec\tau_{\rmn grav}$ and $\vec\tau_{\rmn pres}$, $\vec\tau_{\rmn visc}$ near the sinks does not tend to continuously act in any preferred direction and thus averages out to be much smaller than the gravitational and pressure torques. 

\subsubsection{Binary and multiple formation}

Around 300 yr after the first sink has formed, the disk becomes unstable to fragmentation and a second sink is created. We can estimate the value of $Q$ in the Toomre criterion for disk gravitational instability, 

\begin{equation}
Q = \frac{c_{\rmn s} \kappa}{\pi G \Sigma} < 1  \mbox{\ .}
\end{equation}
 
\noindent Here, $\Sigma$ is the disk surface density and $\kappa$ is the epicyclic frequency, which is equal to the angular velocity for Keplerian rotation. \noindent At this point in the simulation, the soundspeed $c_{\rmn s}$ is approximately 3 km s$^{-1}$ for a $\simeq$ 1000 K disk, and $\kappa$~$\simeq$~$3\times 10^{-11}$~s$^{-1}$ for a disk with Keplerian rotation at the disk outer radius, which we take to be $R_{\rmn disk}$~$\simeq$ 1000 AU.  The disk  mass is close to 30 M$_{\odot}$, and $\Sigma$~$\simeq$~M$_{\rmn disk}/$$\pi R_{\rmn disk}^2$~$\sim$~100~g/cm$^2$.  This gives $Q \sim$~0.4, well-satisfying the fragmentation criterion.

\begin{table}
\begin{tabular}[width=.45\textwidth]{crrrr}
\hline
sink  & t$_{\rmn form}$ [yr] & M$_{\rmn final}$ [M$_{\odot}$]  & r$_{\rmn
init}$ [AU] & r$_{\rmn final}$ [AU]\\
\hline
1  & 0  & 43  & 0 & 0\\
2  & 300 &  13 & 60 & 700\\
3  & 3700  & 1.3 & 930 & 1110\\
4  & 3750 & 0.8 & 740 & 890\\
5  & 4400 &  1.1 & 270 & 240\\
\hline
\end{tabular}
\caption{Formation times, final masses, and distances from the main sink. We include all sinks still present at the
end of the simulation.}
\label{tab1}
\end{table}

\begin{figure*}
\includegraphics[width=.4\textwidth]{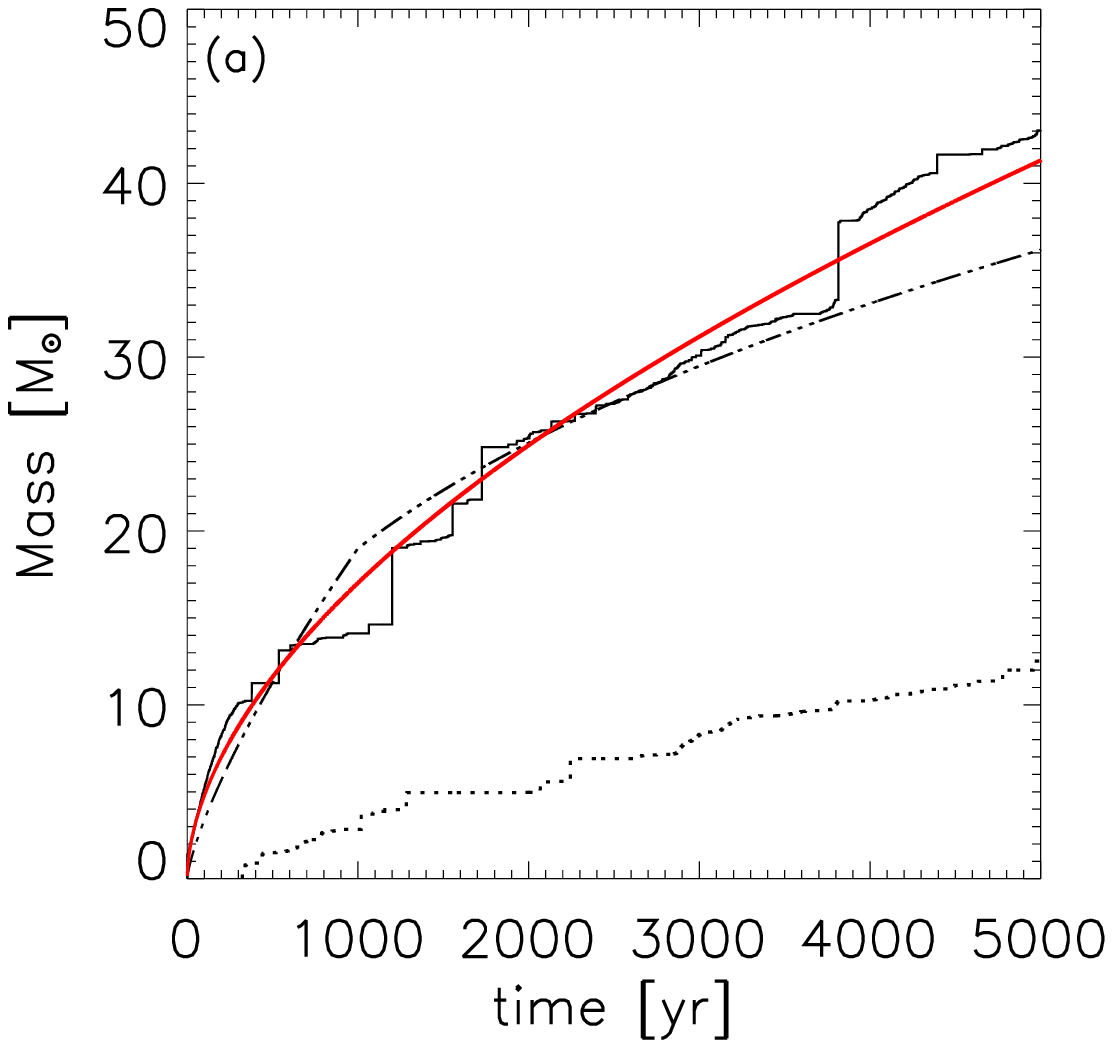}
\includegraphics[width=.4\textwidth]{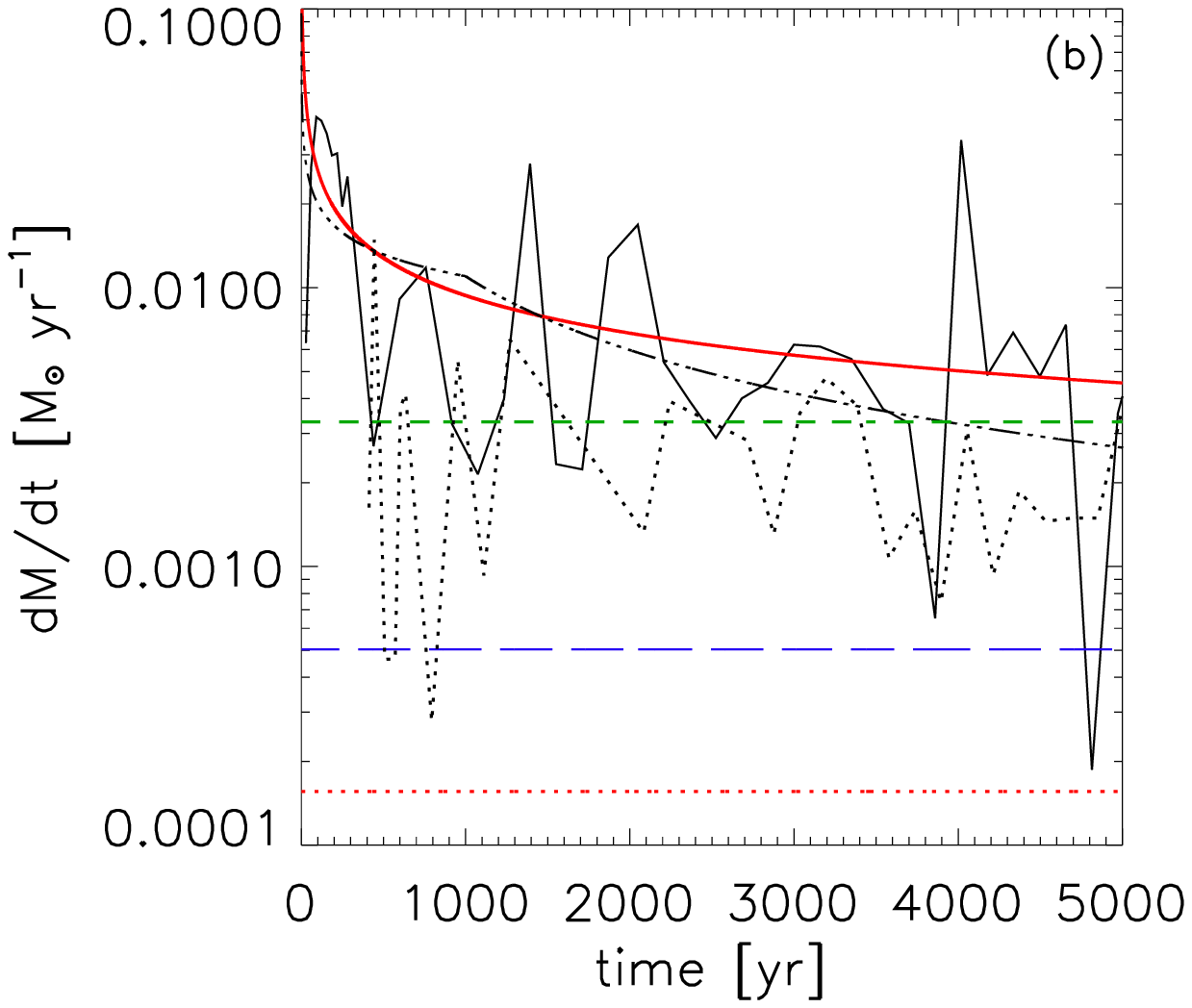}
\caption{{\it(a)} Sink mass versus time since initial sink formation.  The solid line shows the mass of the first sink particle over time in our calculation.   The red line is the least-squares powerlaw fit to the sink mass over time, which goes as  $M\propto t^{0.55}$.  The dash-dotted line shows the result obtained by Bromm \& Loeb (2004). The dotted line is the mass growth of the second largest sink.  {\it(b)}  Accretion rate versus time since initial sink formation.  The solid line shows the accretion rate of the first sink particle throughout our calculation.  The red line is the powerlaw fit to the mass accretion rate ($\dot{M} \propto t^{-0.45}$).  The black dash-dotted line shows the instantaneous accretion rate found by Bromm \& Loeb (2004), while the red dotted line is their average accretion rate over the lifetime of a Pop~III star, determined by extrapolating their rate to 3$\times$10$^6$ yr. The dotted line is the accretion rate of the second largest sink.  Also shown are the Shu accretion rates for isothermal gas at 700 K (green short-dash line) and 200 K (blue long-dash line). }
\label{sinkmass}
\end{figure*}

Soon after the disk instability sets in, a pronounced spiral structure develops, as is evident in Fig.~\ref{morph}. The spiral perturbation serves to transport angular momentum outward through gravitational torques, allowing some of the inner gas to accrete onto the sinks (see \citealt{krumholzetal2007}). By 1000 yr after initial sink formation, several new fragments have formed, represented by new sink particles (shown in Fig.~\ref{morph}).  Some of these fragments eventually merge with other more massive sinks, but by the end of our simulation we have 5 sink particles throughout the disk, described in Table~1.   The two largest sinks are 700 AU apart, have a mass ratio of $\sim$~1/3, and form 300 yr apart (see Sec. 4.2.4 and Fig.~\ref{sinkmass}).  This is not unsimilar to the two fragments found in \citet{turketal2009}, which had a separation of 800 AU, a mass ratio of $\sim$ 3/5, and formed $\sim$ 60 yr apart.  The three smallest sinks in our calculation are all around 1 M$_{\odot}$ and formed after approximately 4000 yr.  Compared to the two largest sinks, these smaller sinks are quite low-mass and have been present in the simulation for only a short time.  The two largest sinks seem to be the only long-term sinks in the simulation.  

 Analytical estimates concerning massive disk fragmentation by \cite{kratter&matzner2006} showed interesting similarity to our results.  They find that protostellar disks hosting massive stars (O and B stars) are prone to fragmenation for disk radii larger than 150 AU, and they predict that many low-mass companions will form in the disks around these stars at initial separations of 100-200 AU (compare with our Table \ref{tab1}).  Though their estimate is based on optically thick disks undergoing viscous heating as well as stellar irradiation, the stabilizing effect of those heat sources is counteracted by rapid stellar accretion and high disk angular momentum.  If protostellar feedback were included in our work,   
this could reduce the number of fragments that form, since the initial protostar could heat the disk and inhibit further fragmentation, perhaps leaving a binary instead of a mutiple system.  This effect is shown in  \citet{krumholzetal2007}, where the evolution of high-mass prestellar cores in the present-day Universe is simulated. They find that protostellar radiative feedback inhibits fragmentation and that the majority ($\ga$ 90 percent) of accreted stellar mass goes onto one protostar, though a small number of lower-mass stars can form at sufficiently large distances from the main protostar or in disk fragments formed through Toomre instability.  It will thus be important to include protostellar feedback in future simulations to fully ascertain the multiplicity of
Pop~III star formation.

\subsubsection{Accretion rate}
How do the initial hydrostatic cores develop into fully formed Pop~III stars?
Fig.~\ref{sinkmass} shows the growth of the two most massive sink particles over time.   We find that the main sink approximately grows as  $M \propto t^{0.55}$ (red line).  By 5000 yr after sink formation, the main sink has grown to 43 M$_{\odot}$, only slightly higher than the corresponding result from \cite{bromm&loeb2004}. The jumps in mass are instances where smaller sinks merged with the largest sink.  One caveat to keep in mind is that we did not explicitly check that these smaller sinks were not centrifugally supported against infall towards the main sink before they were merged.  However, we examined the largest secondary sink ($\sim$ 4  M$_{\odot}$) that merged with the main sink.  This merger occured at 3700 yr after initial sink formation.  We checked the secondary sink's specific angular momentum at the last simulation output before it merged with the main sink.  At this time it was $\sim 100$\,AU away from the main sink, and its specific angular momentum with respect to the main sink was 2$\times$10$^{21}$ cm$^2$ s$^{-1}$.  This is less than the specific angular momentum required at this distance for centrifugal support against infall, 2.6$\times$10$^{21}$ cm$^2$ s$^{-1}$. It is thus plausible that this was indeed a true merger.      

In Fig.~\ref{sinkmass}, we also show the corresponding accretion rate onto the most massive sink particle over time (panel b).  The accretion rate is highly variable and initially very large: $\dot{M}\sim$~7c$_{s}^{3}/G$, if we evaluate the sound speed at 700\,K, and $\sim$~50c$_{s}^{3}/G$ at 200\,K.  These rates are similar to those predicted in analytic solutions (\citealt{larson1969,penston1969,shu1977}), and to the result found in \cite{bromm&loeb2004}.    
Note that the accretion rate does generally decline over time approximately as $\dot{ M}$ $\propto t^{-0.45}$, except for a few `spikes' in accretion when the main sink merges with other smaller sinks.  For comparison the accretion rate onto the second largest sink (dotted line) is also shown, whose powerlaw fit goes as $\dot{ M}\propto t^{-0.39}$ (fit not shown) and is normalized to a value of about 1/5 that of the main sink.  The lower rate is expected since the second largest sink accretes significantly cooler gas.    

As accretion onto the main sink continues, the overall average accretion rate will drop to a much lower value, closer to the average value found in \cite{bromm&loeb2004}.  If accretion were to continue for the entire lifetime of the star, 3 $\times$ 10$^6$ Myr, and if the accretion rate continued to decrease as approximately $\dot{ M}$ $\propto$ t$^{-0.45}$, the final Pop~III star would reach $M_{\ast}\la$~1000~M$_{\odot}$.  However, our fit likely overestimates the accretion rate at late times, and the actual final mass of the star is likely to be significantly lower. Furthermore, feedback effects will likely slow accretion, or even halt it entirely, before the star dies. Thus, our extrapolation serves as a robust upper limit to the final Pop~III stellar mass.  

\subsubsection{Thermodynamics of accretion flow}
\begin{figure*}
\includegraphics[width=.7\textwidth]{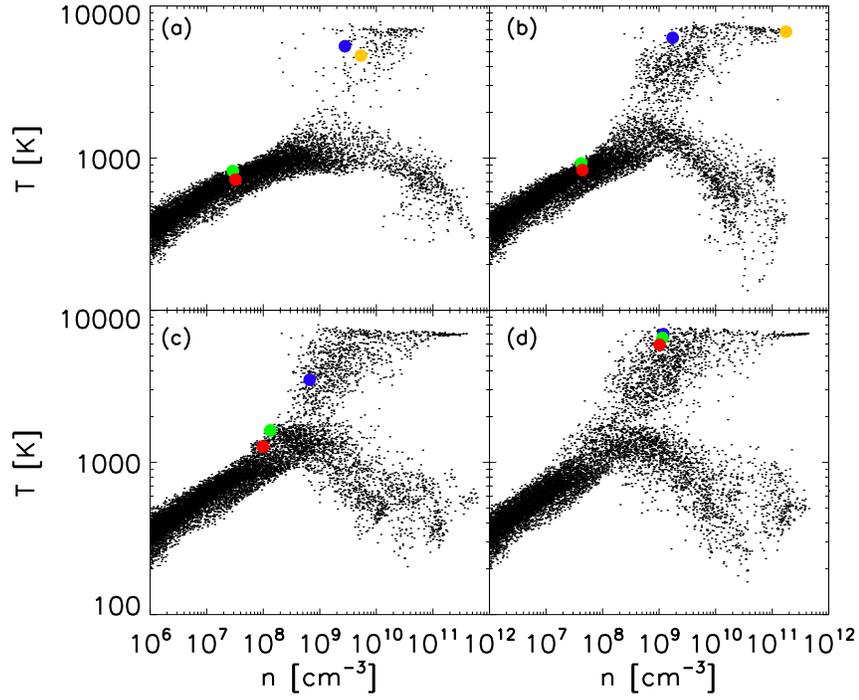}
\caption{Evolution of thermodynamic properties. We plot temperature vs. number density at various times during the simulation. The colored dots trace 
the position of four representative particles, illustrating how the hot phase is populated. Note the hot particle that is accreted by the main sink $\
sim 2000$\,yr after the sink's creation (yellow dot). The times are as follows: {\it(a)} 1000 yr. {\it(b)} 2000 yr. {\it(c)} 3000 yr. {\it(d)} 5000 yr.}
\label{temp_vs_nh}
\end{figure*}

After the initial sink particle has grown in mass, the surrounding gas divides into two phases - a hot and cold one. In Fig.~\ref{temp_vs_nh}, we illustrate this bifurcation with a temperature-density phase diagram at various stages of the simulation. Heating becomes significant once the initial sink grows beyond 10 M$_{\odot}$.  At this mass the gravitational force of the sink particle is strong enough to pull gas towards it with velocites sufficiently high to heat the gas to a maximum temperature of $\sim$ 7,000\,K (see discussion in Sec. 4.2.1).  This is the temperature where the collisional excitation cooling of atomic hydrogen begins to dominate over the adiabatic and viscous heating.  The increasing mass of the sink causes a pressure wave to propagate outward from the sink and heat particles at progressively larger radii and lower density, as is visible in  Fig.~\ref{morph} and Fig.~\ref{temp_vs_nh}. The heated region extends out to $\sim$~2,000\,AU at 5000\,yr. The pressure wave thus propagates at a subsonic speed of $\sim$ 2~km~s$^{-1}$.   

\begin{figure*}
\includegraphics[width=.4\textwidth]{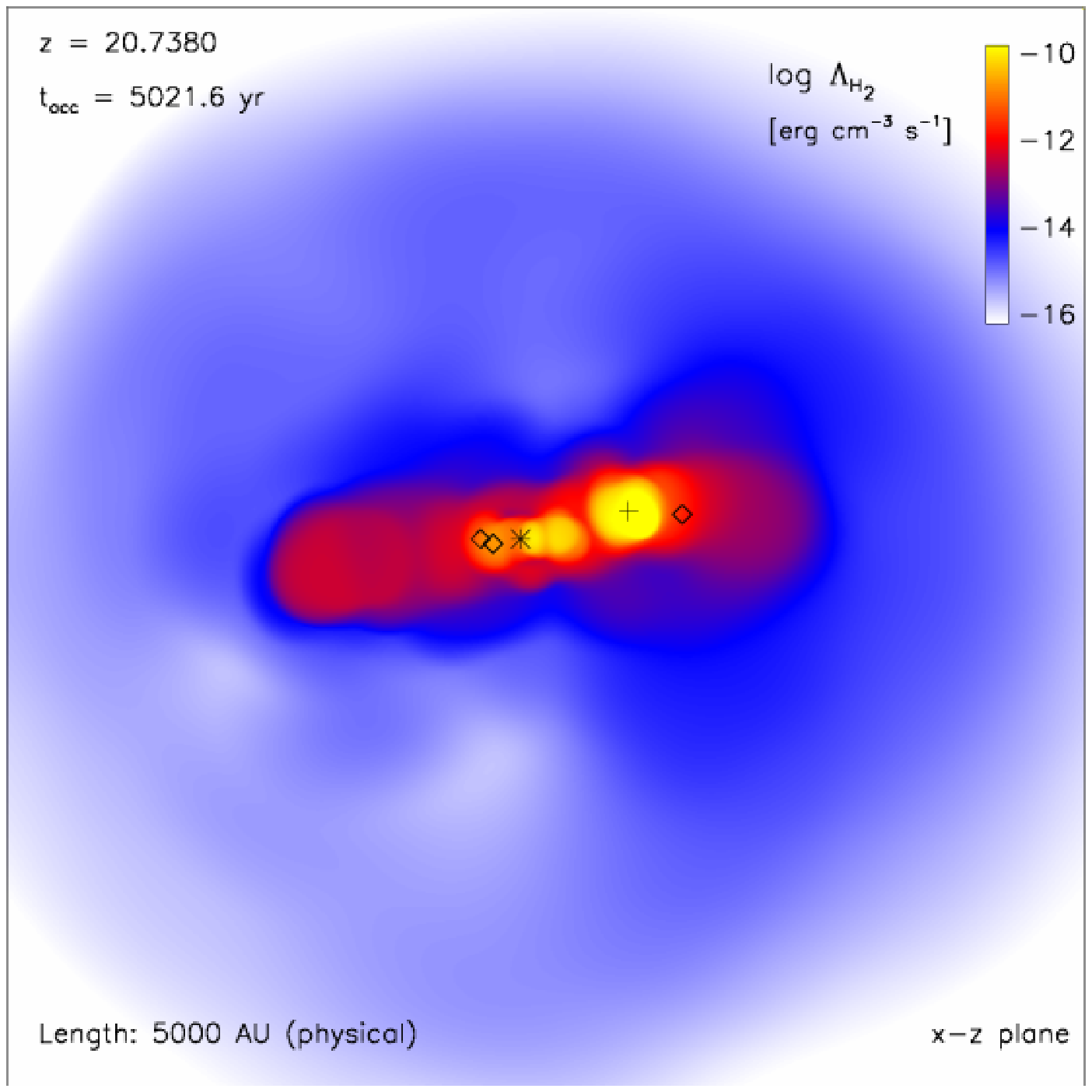}
\includegraphics[width=.4\textwidth]{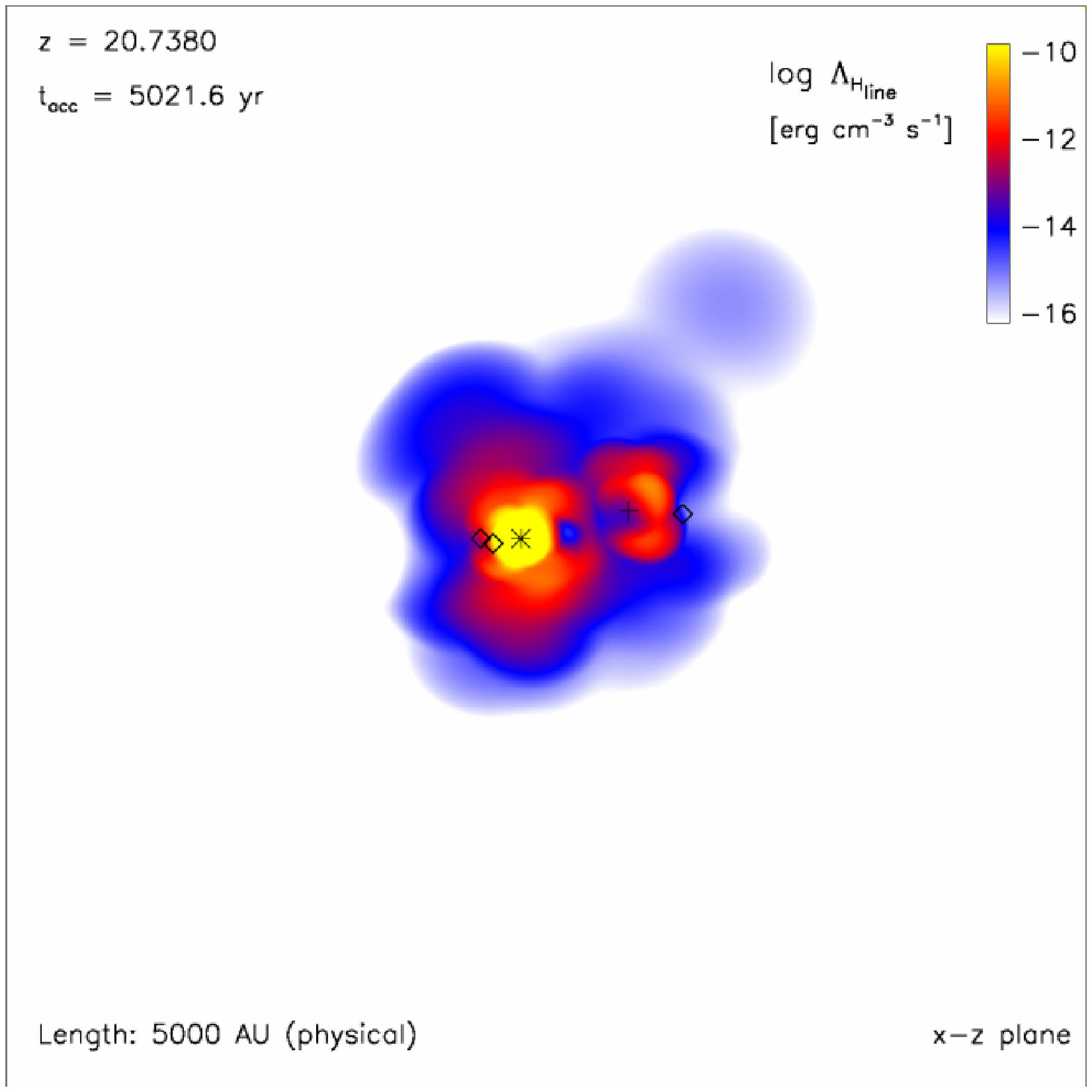}
\caption{Dominant cooling processes in the center of the minihalo.  Sinks are denoted as in Fig.~\ref{morph}. $\itl{Left:}$ H$_2$ cooling rate within the central 5000\,AU, shown in projection along the x-z plane, after $\sim$ 5000\,yr of accretion.  $\itl{Right:}$  H line cooling rate in the same region and at the same time, again shown in projection along the x-z plane.}
\label{cooling_morph}
\end{figure*}

The main sink is able to accrete a fraction of these heated particles (e.g. the yellow particle in Fig.~\ref{temp_vs_nh}), while it continues to accrete cold particles from the disk as well. We record the temperatures of the SPH particles accreted onto the sink, so that we can track the relative contributions from the two phases. When the main sink has grown to $\sim 30$ M$_{\odot}$, accretion from the hot phase begins to occur, such that heated particles contribute slightly over 50\% to the total rate towards the end of the simulation.  By this time, the average temperature of all particles accreted onto the main sink is close to 3000\,K, a value indicative of significant contributions from the hot phase.  The second-most massive sink, on the other hand, accretes particles from only the cool phase as it has not yet reached a sufficiently high mass to capture heated particles.  It has, however, just become massive enough to begin heating its surrounding particles.
Fig.~\ref{cooling_morph} reveals the location of the two different gas phases in a complementary way by showing where cooling due to H$_2$ (left panel) and atomic hydrogen line cooling dominate (right panel). H$_2$ cooling is important in the cold, dense gas of the disk-like structure, whereas atomic cooling occurs in the hot, nearly spherical region around the largest sink particle. This region grows in size as the sink itself gains mass. By the end of the simulation, a similar heated region is also beginning to expand around the second-most massive sink.

Since the heated region is created by the sink particle's potential well, we can estimate its extent by evaluating the Bondi radius, which marks the distance out to where the gravitational energy of the sink will be greater than the thermal energy of the gas. If we use the final mass of the main sink, $M_{\rmn sink} \simeq$ 40 M$_{\odot}$, and the maximum soundspeed of the gas particles, $c_{\rmn s} \simeq 7$ km s$^{-1}$ for temperatures of 7000 K, we find 

\begin{equation}
R_{\rmn B} \simeq \frac{G M_{\rmn sink}}{c_{\rmn s}^2} \simeq 750 \rmn AU \mbox{\ .}
\end{equation} 

\noindent This is  close to the size of the heated region seen in both Fig.~\ref{morph} and Fig.~\ref{cooling_morph}.  We should further note that, except for sink masses lower than $\sim$ 2 M$_{\odot}$,  $R_{\rmn B}$ will be larger than $r_{\rmn acc}$, which is held constant at 50\,AU. When the main sink reaches larger masses and begins to accrete heated particles,  $R_{\rmn B}$ is in fact over an order of magnitude larger than $r_{\rmn acc}$, indicating that our treatment of accretion is indeed conservative and that there is little un-physical accretion of heated particles.

\subsubsection{Feedback}

As noted earlier, the formation of a disk-like configuration around the protostar will have important implications for feedback effects.  As demonstrated by \cite{mckee&tan2008}, the disk structure will mitigate the impact of protostellar feedback. Radiation will be able to escape along the polar directions, and although the innermost, optically thick, part of the accretion disk is not resolved in our simulation, it would shield a portion of the outer accretion flow from direct feedback. Future three-dimensional simulations that include radiative feedback will yield a better understanding of how and when disk accretion onto a primordial protostar may be terminated.
We now wish to assess how important this neglected feedback would be
up to the stage simulated here.

\begin{figure}
\includegraphics[width=.45\textwidth]{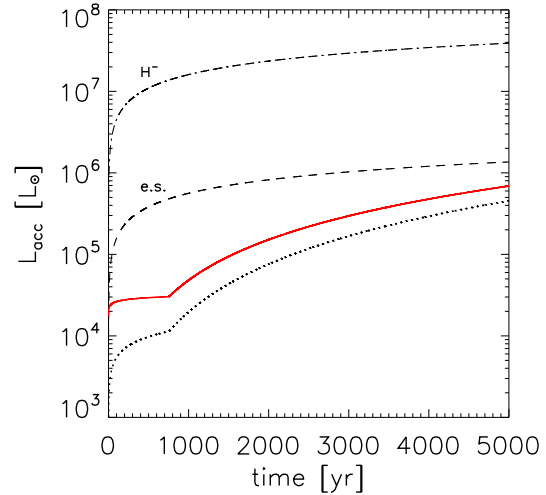}
\caption{Impact of feedback on the accretion process. Solid red line is the accretion luminosity $L_{\rmn acc}$ of the most massive sink versus time, calculated using $R_* = R_{*I}$ at early times and $R_{*} = R_{*II}$ at later times, as described in the text.  In determining $L_{\rmn acc}$ we use the power-law fit to the accretion rate ($\dot{M}$ $\propto$ $t^{-0.45}$).  Dotted line is the prediction of  $L_{\rmn acc}$ by Tan \& Mckee (2004) given $f_{\rmn d}$ = 1 and  $K'$ = 1.33.  Dashed line is the Eddington luminosity $L_{\rmn Edd}$ due to electron scattering, calculated using the sink mass. The dash - doted line is $L_{\rmn Edd}$ due to H$^-$ bound-free opacity, calculated at a temperature of 6000 K. Note that the kink in the curves for the accretion luminosity is an artefact of our idealized modelling of protostellar structure.}
\label{Lacc}
\end{figure}

In Fig.~\ref{Lacc}, we compare the accretion luminosity, $L_{\rmn acc}$, with the Eddington luminosity due to electron scattering and due to H$^-$ opacity.  Determining  $L_{\rmn acc}$ requires an estimate of the protostellar radius $R_*$.  In the initial adiabatic accretion phase, before Kelvin-Helmholtz contraction has commenced, the photospheric opacity of the protostar is dominated by H$^-$ bound-free absorption (i.e., `Phase I' of \citealt{omukai&palla2003}).  The extremely strong sensitivity of the H$^-$ bound-free opacity to temperature ($\kappa_{\rmn H^-} \propto T^{14.5}$) locks the photospheric temperature to $\sim$~6000~K.  We estimate that the protostar emits as a blackbody at this temperature and furthermore assume that the protostellar luminosity during this phase is dominated by $L_{\rmn acc}$. This is justified as long as $t_{\rmn acc}\la  t_{\rmn KH}$, where $t_{\rmn acc}$ is the accretion timescale and $t_{\rmn KH}$ the Kelvin-Helmholtz time. We then have

\begin{equation}
L_{*I} \simeq L_{\rmn acc} = \frac{G M_* \dot{ M}}{R_*} = 4 \pi R_*^2 \sigma_{\rmn SB} T^4   \mbox{\ ,}
\end{equation}

\noindent where $L_{*I}$ is the protostellar luminosity during the adiabatic accretion phase, $\sigma_{\rmn SB}$ is the Stefan-Boltzmann constant, and $T=6000$\,K.  This results in

\begin{equation}
R_{*I} \simeq 50 {\rmn R_{\odot}} \left(\frac{M_*}{\rmn M_{\odot}}\right)^{1/3} \left(\frac{\dot{M}}{\dot{M}_{\rmn fid}}\right)^{1/3}   \mbox{\ ,}
\end{equation}
  
\noindent  where $R_{ *I}$ is the protostellar radius during the adiabatic accretion phase, and $\dot{M}_{\rmn fid}\simeq 4.4\times 10^{-3}$~M$_{\odot}$\,yr$^{-1}$, the fiducial accretion rate used by \citet{stahler&palla1986} and \citet{omukai&palla2003}. Our simple estimate of how $R_{*I}$ varies with mass and accretion rate is quite similar to the results of these more detailed previous studies,
 and here serves to highlight the relevant physics.

During the next phase, the protostellar radius begins to shrink due to Kelvin-Helmholtz contraction. The major source of opacity in the interior of the protostar is electron scattering, which is independent of density and temperature.  This along with hydrostatic equilibrium yields the mass-luminosity relation $L_* \propto M_*^3$.  Using figure 4 of \citet{omukai&palla2003}, we normalize this relation as follows:

\begin{equation}
L_{ *II} \simeq 10^4 {\rmn L_{\odot}} \left(\frac{ M_*}{10 ~ \rmn M_{\odot}}\right)^3 \mbox{\ ,}
\end{equation}

\noindent where $L_{*II}$ is the protostellar luminosity during the Kelvin-Helmholtz contraction phase.  If we further assume that $t_{\rmn KH}$ $\simeq$ $t_{\rmn acc}$, we arrive at

\begin{equation}
\frac{ GM_*^2}{R_* L_*} \simeq \frac{M_*}{\dot{M}} \mbox{\ .}
\end{equation}

\noindent Using $L_{*}\simeq L_{ *II}$, we now find a relation for $R_{ *II}$, the protostellar radius during the Kelvin-Helmholtz contraction phase:

\begin{equation}
R_{*II} \simeq 140 {\rmn R_{\odot}} \left(\frac{\dot{M}}{\dot{M}_{\rmn fid}}\right) \left(\frac{M_*}{10 \rmn M_{\odot}}\right)^{-2} \mbox{\ .}
\end{equation}

In determining the evolution of $L_{\rmn acc}$, we set $R_*=R_{*I}$ at early times when the protostar is still experiencing adiabatic accretion.  When $M_*\sim~10$\,M$_{\odot}$, after approximately 800\,yr of accretion, the value of $R_{*II}$ falls below that of $R_{*I}$.  At this point, we estimate that Kelvin-Helmholtz contraction has commenced and switch to setting $R_*=R_{*II}$.  The resulting transition in the growth of $L_{\rmn acc}$ can be seen in Fig.~\ref{Lacc}.  
The discontinuous slope in the accretion luminosity is an artefact of our
approximate modelling of protostellar structure.

Since the accretion luminosity does not exceed either Eddington luminosity, we conclude that for the duration of our simulation, radiation pressure is not likely to significantly change the early accretion rates we have found. At later times, when the protostar reaches higher masses, however, and in particular when it reaches the ZAMS and hydrogen burning turns on, such feedback effects will become more important, and can no longer be neglected.

\section{Parameter Sensitivity}
\begin{figure}\includegraphics[width=.45\textwidth]{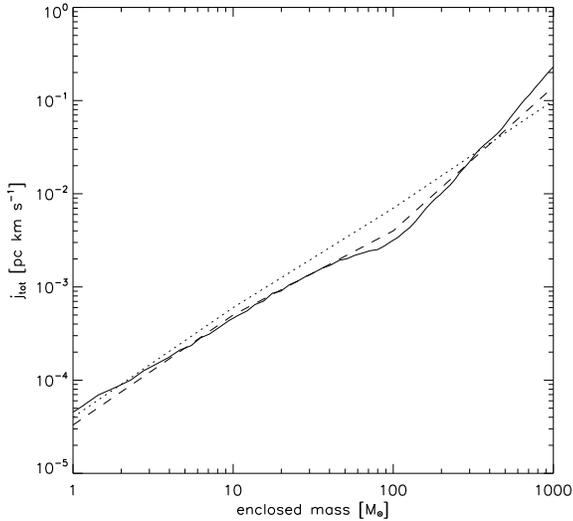}
\caption{Comparison of specific angular momentum profiles for three different cosmological simulations: this work at the point immediately before initial sink formation (solid line), Yoshida et al. (2006, dotted line), and Abel et al. (2002, dashed line).  Despite the differing initial conditions for each simulation, the initial angular momentum distributions are quite similar, particularly at the smallest enclosed mass scales from which most of the mass of the sinks are accreted.}
\label{am_comp}
\end{figure}

 How commonly do Pop III stars form in multiples?  To what extent does this result depend on the initial cosmological parameter values, particularly the value for $\sigma_8$?  Is  there a relation between the spin of a minihalo and the probability that it will host a Pop III multiple (see section 4.1)?  To address this uncertainty, we make an initial estimate of the sensitivity of our results to differing values of  $\sigma_8$ by comparing the specific angular momentum profile of the central 1000 M$_{\odot}$ of our simulation with that of \cite{abeletal2002} and \cite{yoshidaetal2006} (Fig. \ref{am_comp}). The angular momentum profile for our work is taken at the point immediately before the first sink forms.   \cite{abeletal2002} initialized their simulation using h=0.5, $\Omega_{\Lambda}=0$, $\Omega_{\rmn B}=0.06$ and $\sigma_8=0.7$.  \cite{yoshidaetal2006} used  h=0.7, $\Omega_{\Lambda}=0.7$, $\Omega_{\rmn B}=0.04$ and $\sigma_8=0.9$.  If the particular parameter values used in initializing our calculation made the central gas more prone to develop a flattened disk-like structure that would later fragment, this would first be apparent in an enhanced angular momentum profile.  However, despite the varying initial conditions used for each of these calculations, the angular momentum profiles are all quite similar in shape and amplitude, especially in the innermost region.  The central 100 M$_{\odot}$ that will later become the star-disk system in our calculation does not differ in angular momentum from the other two calculations by more than a factor of 2 to 3.  In fact, in this mass range the simulation of \cite{yoshidaetal2006} tends to have the highest angular momentum.  The initial angular momentum distribution of the star-forming gas thus seems fairly robust within the range of cosmological parameter values sampled with these three simulations, and our overall result should be correspondingly robust.  We further note that \cite{clarketal2008} find that their non-cosmological simulation of a primordial gas cloud also resulted in a profile very similar to the same two cosmological simulations, and upon extending their calculation through the sink particle method they also found extensive fragmentation.          

However, the generality of our result will be best determined through a more comprehensive set of simulations, aiming at studying fragmentation over a range of minihalo spins. We will report on the results in a future work.   Finally, we expect that the effects from protostellar feedback, neglected here, will likely dominate over those arising from variations in the initial conditions.  Again, this needs to be addressed with improved simulations. 

\section{Summary and Discussion}
We have evolved a cosmological simulation until a primordial minihalo has formed at $z\simeq 20$, and follow the subsequent collapse of the gas in its center up to densities of $n=10^{12}$\,cm$^{-3}$.  We find that the initial collapse, leading to the formation of a small hydrostatic core, is very similar to previous high-resolution work. Subsequently, a disk-like structure grows around the central core, eventually comprising a mass of $\sim$~40~M$_{\odot}$. We use the sink particle method to study the mass flow onto the central protostar for 5000\,yr after its formation.  Similar to the fragmentation found in \citet{turketal2009}, we find that a second protostar forms soon after the first one.  Upon further evolution, the disk exhibits spiral structure and undergoes further fragmentation, resulting in several more sink particles by the end of the simulation.  Most of the mass of the smaller, secondary sinks is accreted through the disk, which consists mostly of cold gas. Once the main sink grows beyond 10\,M$_{\odot}$, its strong gravity is able to heat the surrounding particles to several thousand Kelvin, such that a two-phase medium emerges. The main sink is able to accrete these heated particles as well. After 5000\,yr of accretion, the most massive sink has reached $\sim$\,40~M$_{\odot}$, at which point we terminate our simulation. Beyond this time, protostellar feedback effects, which are not included here, could no longer be neglected. The Pop~III star, however, is expected to accumulate more mass. Even in the no-feedback case, however, the finite time available for accretion limits the Pop~III mass to $\la 1000$\,M${_\odot}$, where it is evident that in realistic settings, masses will not reach such extreme values.  The accretion rate and the number of stars that form will be affected by protostellar feedback, with the likely outcome that both individual masses and
the number of stars will be reduced (see \citealt{krumholzetal2007}).

The recent radiation-hydrodynamical simulation by \cite{krumholzetal2009}, which studies the collapse of a massive pre-stellar core in a present-day GMC, shows interesting similarity to our investigation of the primordial case. They followed the collapse of a 100\,M$_{\odot}$ gas cloud, the formation of several protostellar seeds and their subsequent growth by accretion. Like our simulation, their system also evolved into a disk with a radius of order 1000\,AU which fragmented due to gravitational instability. Similar to the sink mergers seen in our calculation, in their simulation the secondary stars that formed in the disk collided with the most massive star. By the end of their simulation, which was halted after $\sim$~60,000 yr, the system had evolved into a binary with stars of mass $\sim$~30 and 40~M$_{\odot}$, masses fairly close to those of the two largest sinks in our simulation.  Their calculation, however, included dust opacity as well as radiation pressure from the stars, and their initial conditions were typical of present-day star formation.  Their accretion and fragmentation timescales were also much longer than those found in our calculation, but the overall qualitative results were alike.

Thus, the formation of binary and multiple systems within disk structures is likely not limited to modern-day star formation and can be found even in the primordial case, arising from cosmological initial conditions.   Binary and multiple systems may be much more common in the Pop III case than previously thought.  
 As similarly argued in \cite{clarketal2008}, 
this is because most cosmological simulations do not follow the gas evolution for many years after the first protostar has formed, and fragmentation may not typically occur until after the simulations are stopped.   For instance, the second protostar in our calculation did not form until 300\,yr after the first, while the calculation of \cite{turketal2009} was stopped only 200\,yr after the first protostar formed.  

If Pop III binaries and multiples are actually common, then this has important implications for the final fate and observational signature of Pop~III.1 stars, those zero-metallicity stars that have not yet been influenced by any previous stellar generation (e.g. Bromm et al. 2009).  The presence of multiple accretors in our simulation did not seem to affect the accretion rate onto the largest sink as compared to previous work (Bromm \& Loeb 2004).  However, extending our simulation to later times might show that the most massive star would ultimately have to compete with its secondary companion for mass.  As pointed out in \citet{turketal2009}, if this in the end prevents the largest star from reaching very high masses ($\ga$ 100 M$_{\odot}$), this would also affect its resulting chemical signature. If multiple star formation were common in Pop~III, then the occurrence of pair-instability supernovae (PISNe) might be significantly reduced, since stars that explode as PISNe must be in the higher mass range (140~M$_{\odot}$~$<$~$M_{*}$~$<$~260~M$_{\odot}$, \citealt{heger&woosley2002}).  Extremely metal-poor Galactic halo stars are believed to preserve the nucleosynthetic pattern of the first SNe, so this may help to explain the lack of PISNe chemical signatures found in halo stars (e.g. \citealt{christliebetal2002,beers&christlieb2005,frebeletal2005,tumlinson2006,karlsson2008}).  \citet{turketal2009} also note that if multiple high-mass Pop~III stars can form within a minihalo at similar times, the ionizing flux on neighboring minihaloes could be increased, possibly leading to more efficient molecular cooling and ultimately lower-mass (Pop~III.2) stars in the neighboring haloes (e.g. \citealt{osheaetal2005}). This again would influence the chemical signature of extremely metal-poor Galactic halo stars.

It should be mentioned that tight binaries on scales smaller than the resolution of our simulation could also form.  Though a higher resolution 
study would be necessary to confirm this, the formation of a fragmenting disk of steadily increasing angular momentum in our cosmological simulation makes this a reasonable possibility.  This could also limit the mass of Pop~III stars, yielding similar effects on their resulting chemical signature as discussed above.  The possible binary nature of Pop~III stars, provided that they would not grow to become very massive, may also yield a direct primordial formation pathway to carbon-enhanced metal-poor (CEMP) stars, those metal-poor halo stars with unusually high carbon abundance (e.g. \citealt{frebeletal2007,tumlinson2007}).  If one of the members of a Pop~III binary was an intermediate-mass star (1~M$_{\odot}$~$\la$~$M_{*}$~$\la$ 8 M$_{\odot}$), then the companion might undergo carbon enhancement through binary mass-transfer (e.g. Suda et al. 2004).  This would occur when the intermediate-mass star enters an asymptotic giant branch (AGB) phase during which its C-rich ejecta can be captured by the companion.  
Furthermore, another type of Pop III signature would arise if gamma-ray bursts (GRBs) were able to form from very tight Pop III binaries.  These may be detectable by {\it{Swift}} (see \citealt{bromm&loeb2006,belczynskietal2007}).

On the other hand, if both members of a typical Pop~III binary were in the mass range to collapse directly into a black hole ($M_{*}$~$\ga$~260~M$_{\odot}$), then they could become massive black hole (MBH) binaries.  Such a population of MBH binaries may emit a gravitational wave (GW) signature detectable by the future space-born {\it Laser Interferometer Space Antenna (LISA)}. Previous studies have already explored the possibility of GW detection of MBH binaries formed through halo mergers during early structure formation (e.g. \citealt{wyithe&loeb2003,sesanaetal2005}).  \cite{sesanaetal2005} find that high redshift ($z \ga$ 10) MBH binaries with masses $\la$ 10$^3$ M$_{\odot}$ will be detectable by $LISA$.  MBH binaries formed from Pop III progenitors thus approach the range of detectability by $LISA$, especially if the BHs continue to grow in mass through accretion.  Furthermore, \cite{belczynskietal2004} predict that advanced $LIGO$ could also detect the GWs from Pop III MBH binary mergers.  \cite{madau&rees2001} suggested a different type of Pop III MBH signal when they discussed how an accreting primordial MBH could tidally capture a star and become a detectable off-nuclear ultraluminous X-ray source. They conclude, however, that such a non-primordial origin to Pop~III binaries is extremely inefficient.

A modified picture of Pop III star formation is thus emerging in which disk formation, fragmentation, and the resulting binary and mutiple systems will play an important role in determining the mass of Pop~III stars and their influence on later star and galaxy formation. Further studies, particularly ones  which include protostellar feedback, will continue to add to this developing picture.

\section*{Acknowledgements}

AS thanks Jarrett Johnson and Milos Milosavljevic for many helpful discussions.  TG acknowledges support from the Heidelberg Graduate School of Fundamental Physics (HGSFP).  The HGSFP is funded by the Excellence Initiative of the German government (grant number  GSC 129/1).  VB acknowledges support from NSF grant AST-0708795 and NASA ATFP grant NNX08AL43G. The simulations presented here were carried out at the Texas Advanced Computing Center (TACC).

\bibliographystyle{mn2e}
\bibliography{acc12}{}

\label{lastpage}

\end{document}